\documentclass[11pt]{amsart}

\usepackage{mystyle}
\usepackage{graphicx}
\usepackage{subfigure}

\begin{document}
\begin{titlepage}

\begin{center}

\textsc{\Large Research Initiation Project}\\[0.5cm]

{ \huge \center Persistence Diagrams and The Heat Equation Homotopy}\\[0.4cm]

{\large 17 July 2009} 

\end{center}

\vfill
\begin{flushright} \large
Brittany Terese Fasy \\
Dr. Herbert Edelsbrunner, Adviser \\
Dr. Pankaj Agarwal, Committee Member\\
Dr. John Harer, Committee Member
\end{flushright}

\end{titlepage}

\section{Introduction}\label{sec:intro}
Given two functions, we are interested in comparing the respective persistence 
diagrams. The persistence diagram is a set of points in the Cartesian plane 
used to describe births and deaths of homology groups as we iterate through
sublevel sets of a function.
There are many ways that we can compare two diagrams, including 
matching points between the diagrams.  However, finding a meaningful matching
(see \secref{sec:matching}) is a difficult task, especially if we are 
interested in capturing the relationship between the underlying functions.

As stated in the proposal, the goal of this Research Initiation Project
is to gain an intuition
for persistence and homology, as well as to understand the current state of
research in these fields.  I accomplished this goal by investigating the
following problem:
For a $2$-manifold $M$, 
suppose we have two continuous functions $f,g \colon M \to \R$.
We can create the persistence diagrams for $f$ and for $g$.  
If we know the relationship between $f$ and $g$, we can make informed decisions
when matching points in the persistence diagrams.  In particular, we are
interested in the case where $f$ and $g$ are homotopic.  Given a homotopy
between $f$ and $g$, we can create a vineyard of the persistence diagrams.
Then, we use the vines in the vineyard to help make an informed matching of
the points in the persistence diagrams.

In this paper, we discuss two known
methods of matching persistence diagrams, by measuring the bottleneck and the
Wasserstein distances.  Although stability results exist for matching
persistence diagrams by minimizing either the bottleneck or the Wasserstein 
distance, these matchings are made without 
consideration of the underlying functions $f$ and $g$.  If we are able to 
create a continuous deformation of the function $f$ to $g$, then we can use this
additional information to aid in the matching of the points in the
persistence diagrams.  As a result, the matching obtained will be based
on the underlying functions.  We look into alternate way of measuring the 
distance between the persistence diagrams for the functions by assuming that 
there exists a homotopy between the functions.
We create a homotopy that we call the heat equation 
homotopy and measure distances between the persistence diagrams for $f$ and $g$
by using these homotopies to aid in the pairing of points in the persistence
diagrams of $f$ and ~$g$.  Then, we turn to analyzing 
an example of the heat equation homotopy and discuss various interesting 
patterns.

\section{Computational Topology Preliminaries}\label{sec:back}
Observing patterns and features in data sets is a common goal in many
disciplines, including biology.  Extracting the key features
from a noisy data set can be an ambiguous task, and often involves simplifying
and finding the best view of the data.  
Computational topology, and more specifically 
persistent homology, is a tool used for data analysis.
Here, we give a brief review of the necessary background of 
computational topology, but refer you to 
~\cite{newbook},~\cite{survey} and ~\cite{hatcher} for more details.

\subsection{Homology}
Let $X$ be a simplicial complex of dimension $d.$  For $p \in \N$ and 
$p \leq d$, the symbol $X_p$  will denote the power set of all $p$-simplices 
in $X.$  Each set of $X_p$ is called a \emph{$p$-chain}.  The \emph{chain group} 
$C_p$ is defined by the set $X_p$ under the disjoint union, or symmetric 
difference, operation.  This operation can be interpreted as addition modulo 
two.  The group $C_p$ is therefore isomorphic to $\Z_2$ to some non-negative 
integer power.  All algebriac groups in this paper are vector 
spaces over $\Z_2$.  Consider the boundary homomorphism :
$$ \partial_p: C_p \to C_{p-1},$$ 
that maps the $p$-chain $\alpha \in C_p$ to the boundary of $\alpha,$ 
a chain in $C_{p-1}$ ~\cite{hatcher}.  The  $p^{th}$ homology group of $X$, 
denoted $H_p(X)$, is defined as the kernel of $\partial_p$ modulo the image 
of $\partial_{p+1}$:
\begin{center} $H_p(X)=$ Ker($\partial_p$)/Im($\partial_{p+1}$). \end{center}
The kernel of the homomorphism $\partial_p$ is the set of elements
in the domain that are evaluated to zero (the empty set) 
and the image of $\partial_{p+1}$ is the set
of elements of the form $\partial_{p+1}(x)$, where $x$ is in the domain $C_{p+1}$:
\begin{center} 
   Ker($\partial_p)=\{ \alpha \in C_p | \partial_p(\alpha)= \emptyset \}$ and
\end{center}
\begin{center}
    Im($\partial_{p+1})=\{ \alpha \in C_p | \exists \alpha'  \in C_{p+1}  
    \ni \partial_{p+1}(\alpha')=\alpha \}.$ 
\end{center}
The $p^{th}$ Betti number, $\beta_p$, is the rank of the $p^{th}$ homology 
group of $X.$   By definition, the rank of a group is the (smallest) number of 
generators needed to define the group up to isomorphism.  Since we are 
concerned with groups with $\Z_2$
coefficients, the rank uniquely defines the group up to isomorphism.  
For example, the group with
three generators is $\Z_2^3 = \Z_2 \oplus \Z_2 \oplus \Z_2$ and the group
with $n$ generators is $\Z_2^n$.

\subsection{Persistent Homology}\label{ss:persHomology}
Now, we define persistent homology for functions from $\R$ to ~$\R$.
The complete discussion of the extension of these ideas to higher dimensions
is found in ~\cite{newbook} and ~\cite{survey}.  We present a simplified
setting to focus on the relevant concepts, while avoiding the complications
that arise in the general setting.

Suppose $C$ is the graph of $f \colon \R \to \R$. We can think of $f$ as 
the height function on ~$C$.
Now, we characterize the topology of the \emph{sublevel set}
$\R^f_s = f^{-1}((-\infty,s])$, and we monitor how the homology groups change
as $s$ goes from negative infinity to infinity.
The zeroth persistent homology group, denoted $H_0(\R_s^f)$, will change whenever 
$s$ is 
a local maximum or a local minimum of the function ~$f$.
The maxima and the minima are where the Betti numbers as
well as the homology classes change for functions from $\R$ to $\R$. 
A critical point is defined as the values $r \in \R$ where the derivative is 
zero: $\frac{d}{dx}f(r) = 0$.  Then, $s=f(r)$ is called the 
\emph{critical value} at $r$ ~\cite{stewart}.  A \emph{Morse function} defined 
over a subspace of $\R$ is a smooth function, such that no two critical values
share a function value, and the second derivative at each critical value is non-zero.
For a Morse function $f$, the critical points are the set of $r \in \R$ with
$s=f(r)$, such that  
a Betti number changes by only one from $\R^f_{s-\epsilon}$ to ~$\R^f_{s+\epsilon}$
for every sufficiently small value of $\epsilon > 0$.  If the sum 
of the Betti numbers increases, we call $r$ a positive critical point.  If 
the sum decreases, then $r$ is a negative critical point.

\begin{figure}[hbt]
 \vspace*{0.1in}
 \centering
 \centerline{\epsfig{figure=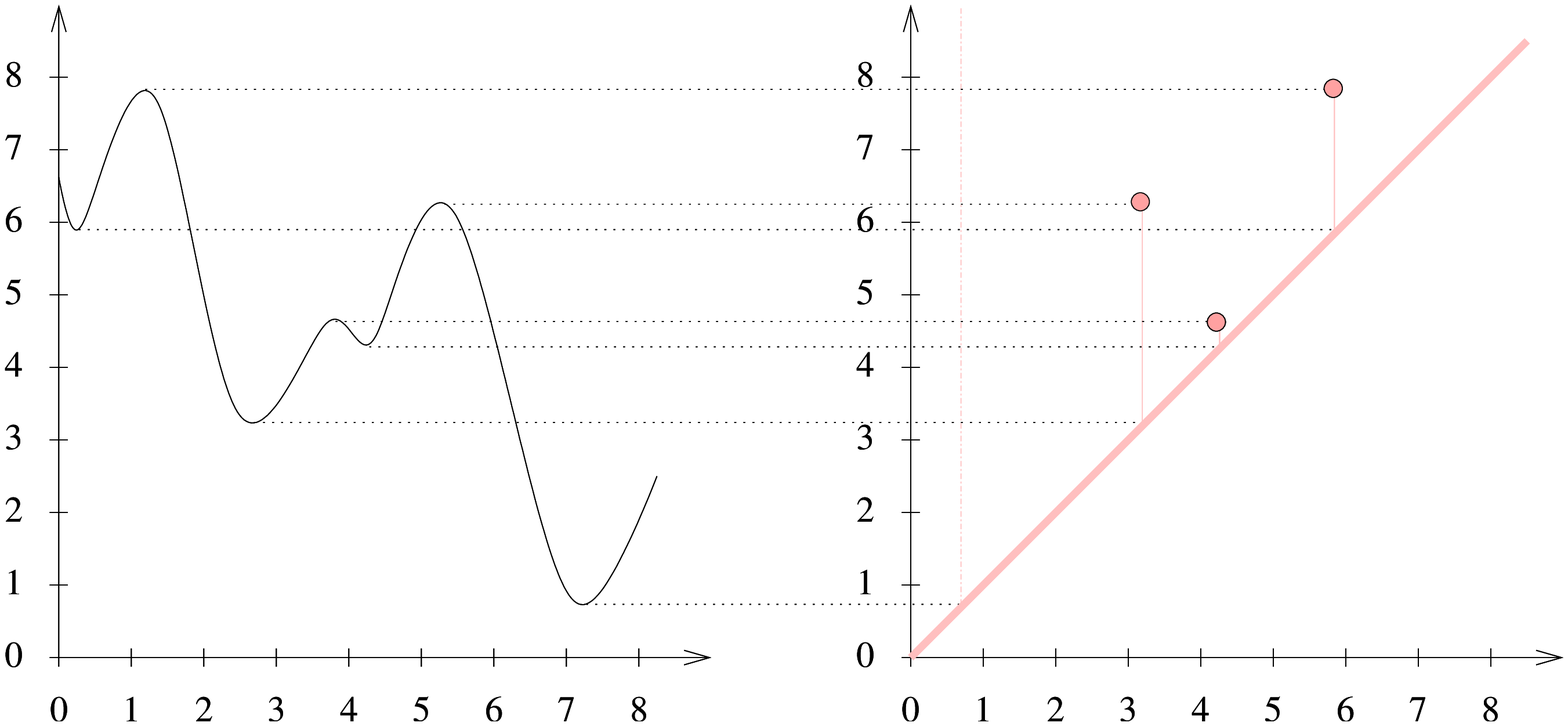,height=2in}}
 \caption{On the left, we see the graph of a function $f$ in ~$\R^2$.  On the right is the
          corresponding persistence diagram, ~Dgm$_0(f)$.  Each point is drawn  
          with multiplicity one.  The birth at the point $(7,1)$ remains unpaired.
         }\label{fig:persistence}
\end{figure}
As $s$ increases from negative infinity, we label each new component with the 
positive critical value that introduces the component. 
We pair each negative critical value $s$ with the most recently discovered 
unpaired positive critical value representing the components joined at $s.$  
This will lead to the diagram Dgm$_0(f)$ as demonstrated in 
\figref{fig:persistence}.  
Consider one pair: a positive critical value that was introduced at time 
$s=s_1$ and a negative critical value that was introduced at time $s=s_2$, where 
$s_1<s_2$.  This pair is represented in Dgm$_0(f)$ as the point $(s_1, s_2).$
The \emph{persistence} of that pair is equal to the difference in function 
values: $s_2-s_1.$  In \figref{fig:persistence}, there are three pairs of 
points and one positive critical value that remains unmatched.  This value
represents an essential homology
class.  It would be paired if we were to consider the extended persistence 
diagram as presented in ~\cite{extended}.

\subsection{Relationships between Diagrams}\label{ss:vineyard}
We now turn to looking at two functions.
Two functions are called homotopic if there exists a 
continuous deformation of the first function into the second.
We are interested in creating a homotopy between
$f$ and $g$ in order to observe how the corresponding persistence diagram 
changes through time.  More importantly, we use the homotopy when
matching points in the persistence diagrams.
Before proceeding, let us formally define a homotopy and give an example
that we will use later.
\begin{definition}[Homotopy]
We say that $f,g \colon \M \to \R$ are \emph{homotopic} if there exists a 
continuous function $F \colon \M \times [0,1] \to \R$ such that $F(x,0)=f(x)$ 
and $F(x,1)=g(x)$, for all ~$x \in \M$.  We will denote the homotopy $F(x,t)$ 
by ~$f_t(x)$.
\end{definition}
\begin{example}[The Straight Line Homotopy]
The straight line homotopy interpolates linearly from each point in the 
continuous function $f$ to the corresponding point in the continuous function 
~$g.$  We can write $f_t(x)=(1-t)\cdot f(x)+t\cdot g(x)$ for all $t \in [0,1]$
and all $x \in M$.
\end{example}

For $f,g \colon \M \to \R$, we have a diagram for each integer 
$p \in \{ 0,1,2\}$.  Now, choose a value of $p$ and assume that we have a 
homotopy (not necessarily the straight line homotopy) from $g$ to ~$f$.  
Choose $\tau$ time-steps of the homotopy, 
$0 < t_1 < \ldots < t_\tau=1.$  Let $t_0=0$, so that Dgm$_p(f_0)$ is the
initial persistence diagram.
At each time $t_j$, we have a persistence diagram Dgm$_p(f_{t_j})$.
Moreover, given Dgm$_p(f_{t_j})$, we can compute
Dgm$_p(f_{t_{j+1}})$ in time linear in the number of simplices of the
filtration by using a straight line homotopy between $f_{t_j}$ and 
$f_{t_{j+1}}$, as described in ~\cite{vines}.  In \secref{ss:vines},
we describe how to use information obtained from this computation in order to pair
points in the persistence diagrams.  Then, we stack the diagrams
so that Dgm$_p(f_t)$ is drawn at height $t$ in $\R^3$
and connect the points in the diagrams by curves formed by the line segments
connecting matched points in consecutive diagrams.  The result is a piecewise
linear path between points in Dgm$_p(f_0)$ and points in Dgm$_p(f_1)$.   
If we let the time 
difference between any two consecutive diagrams approach zero, the piecewise
linear path becomes a set of continuous curves by the stability result for 
the straight line homotopy (see \secref{ss:stability}).
Each curve that traces the path of an off-diagonal points through time 
is called a \emph{vine}.  The
collection of vines is referred to as a \emph{vineyard} ~\cite{vines, newbook}.  
We pair the endpoints of each vine 
to obtain a matching of the persistence points in Dgm$_p(g)$ with the
points in Dgm$_p(f)$.

\section{Matching Persistence Diagrams}\label{sec:matching}
To match the points in Dgm$_p(g)$ and Dgm$_p(f)$ without considering the homotopy, we look 
at methods of finding matchings on bipartite planar graphs.  We begin this 
section with a few definitions.  
 A \emph{matching} is a bipartite graph $P=(A \cup B, E)$ where the
vertex sets $A$ and $B$ are disjoint, the edges are between one vertex
in $A$ and one vertex in $B$, and each vertex is incident on at most one edge.
A matching is \emph{maximal}
if the addition of any edge would result in a graph that is no longer a matching.
A matching is \emph{perfect} if every vertex is incident upon
exactly one edge.  In other words, it is a matching where there does not exist
an unmatched vertex.  

We consider two different methods for measuring the distance
between the persistence diagrams, Dgm$_p(f)$ and Dgm$_p(g)$.  The goal is to 
match every point in $A=$Dgm$_p(f)$ to a point in $B=$Dgm$_p(g)$ in order to minimize 
the cost of the matching.  One method for determining this cost is to 
consider the bipartite matching
problem, where for every $a \in A$ and $b \in B$,
the edge ~$\{a,b\}$ has a cost $C(a,b)$.  Then, the cost of a perfect matching
is the sum (or the maximum) value of the edge costs.  The cost for a matching
that is not perfect is infinite.  
To resolve the issue where the number of off-diagonal points in both diagrams 
is not equal or the diagrams are dissimilar, we allow an off-diagonal point to
be matched to a point on the line $y=x$.
The use of the bottleneck and Wasserstein matchings for this purpose is 
presented in Chapter VIII of ~\cite{newbook}.  As we will show, each of these 
methods entails some notion of stability for the persistence diagrams.
That is, we can bound the bottleneck and Wasserstein distance between 
persistence diagrams by the distance between $f$ and ~$g$, for some class
of functions.

\subsection{Two Cost Metrics}
Consider the matching problem where we must match the elements of set 
$A \subseteq \R^2$ with elements of a second set $B \subseteq \R^2$,  and 
where there is a cost associated with each pair $(a,b) \in A \times B$.  
We will define both the bottleneck
and the Wasserstein costs of a matching.  Later, we will use these
distance metrics as the costs to find the bottleneck and the Wasserstein matchings.
 Let $C(a,b)$ be the $L_\infty$
 distance between the points $a$ and $b$; that is,
 $C(a,b) = \max\{ |a_x-b_x|, |a_y-b_y| \}.$
\begin{definition}[Bottleneck Cost] 
 The bottleneck cost of a perfect matching $P$ is the maximum edge cost:
 $$\max_{(a,b) \in P} C(a,b).$$
\end{definition}
\begin{definition}[Wasserstein Cost]
 The \emph{degree $q$ Wasserstein Cost} of a perfect matching is sum of the
 edge cost over all edges in the matching:
 $$\left( \sum_{(a,b) \in P}{C(a,b)^q} \right)^{1/q}.$$
\end{definition}

\subsection{The Bottleneck Matching Criterion}\label{ss:bottleneck}
The bottleneck matching minimizes the bottleneck cost of a matching 
 over all perfect matchings of $A$ and ~$B$:
$$ W_{\infty}(A,B) = \min_{P} \max_{(a,b) \in G} C(a,b).$$
The use of the notation $W_{\infty}$ to denote the bottleneck distance
will become clear in the next section.

If $|A| = |B| = n$, then a maximal matching can be found in $O(n^{5/2})$ 
using the Hopcroft-Karp algorithm ~\cite{hopKarp}.  If we 
mimic the thresholding approach of the hungarian method ~\cite{hungarian}, then
the bottleneck solution can be found in ~$O(n^{5/2}\log n)$.
Since $A$ and $B$ are sets of points in the plane, we can improve the 
computational complexity of determining the matching under the bottleneck
distance.  Efrat, Itai, and Katz developed a geometric improvement to the 
Hopcroft-Karp algorithm with a running time of $O(n^{1.5}\log^2 n)$ 
~\cite{Efrat01geometryhelps}.

\subsection{The Wasserstein Matching Criterion}\label{ss:wmatch}
In the Wasserstein Matching, we seek to minimize the maximum 
degree $q$ Wasserstein cost over all perfect matchings:
$$W_{q}(A,B) = \min_{P} \left( \sum_{(a,b) \in M}{C(a,b)^q} \right)^{1/q}.$$
Although for small values of $q$, the bottleneck and the Wasserstein criteria
may produce different matchings, if we take the limit as $q \to \infty$, we see
that the Wasserstein criterion approaches the bottleneck criterion.

The Hungarian method computes the Wasserstein matching in $O(n^4)$ computational
complexity ~\cite{hungarian}.  In ~\cite{vaidya}, Vaidya maintains weighted
Voronoi diagrams for an 
$O(n^{2.5}\log n)$ computation of this matching.  Further improvements were made
by Agarwal, Efrat, and Sharir ~\cite{agarwal1995}. They 
utilize a data structure that improves the running time to $O(n^{2+\epsilon})$ 
for the min-weight Euclidean matching.

\subsection{Stability Theorems}\label{ss:stability}
A distance metric (and corresponding matching) is stable if a small change in
the input sets $A$ and $B$ produces a small change in the measured distance
between the sets.  The property of stability is not obvious and sometimes 
not true.  If a matching is stable, however, we can use it to create a
vineyard from a smooth homotopy.

Let $P_{\infty}$ be the matching obtained using the bottleneck criteria; that is, the 
matching $P_{\infty}$ is the matching that minimizes the bottleneck distance.
Let $f,g$ be tame functions.  This means that the homology groups of $\M_s^f$ and 
of $\M_s^g$ have finite rank for all $s.$  In addition, only a finite number of 
homology groups are realized as $H_p(\M_s^f)$ or $H_p(\M_s^g)$ ~\cite{survey}.
The function difference $||f-g||_\infty$ is the maximum difference between the 
function values:  $$ ||f-g||_\infty = \sup_{x \in X} | f(x)-g(x) |. $$
The Stability Theorem for Tame Functions, which gives us that the bottleneck 
distance $W_{\infty}(\text{Dgm}_p(f), \text{Dgm}_p(g))$ is bounded above by 
$||f-g||_\infty$ ~\cite{cohen2007stability}.  
The proof of this theorem 
presented in \cite{vines} uses the following result:
\begin{result}{Stability Result of the Straight Line Homotopy}
 Given $f_t(x)$, the straight line homotopy from $g$ to $f$, we know that
there exists a perfect matching $P$ of the persistence diagrams for $f$ and $g$ 
such that the bottleneck cost of $P$ is upper bounded by the distance between $f$ and $g$:
$$ \max_{(a,b) \in P} C(a,b)  \leq ||f-g||_{\infty}. $$
\end{result}
We will explain how to obtain this matching $P$ in \secref{ss:hehoMatching}.

The Wasserstein distance is stable for
Lipschitz functions with bounded degree $k$ total persistence.  
This is proven in ~\cite{yuriy}.  If we relax
either of these two conditions, then 
the Wasserstein distance becomes unstable for two functions $f$ and $g$ where 
$||f-g||_{\infty} \leq \epsilon$ as $\epsilon$ approaches zero.

\section{The Heat Equation}\label{sec:dist2}
The heat equation is a mathematical description of the dispersion of heat 
through a region in space.   We start with the initial reading (or a guess)
of the temperature throughout an enclosed space, say an empty room.
After infinite time and given no external changes,
the temperature at each point in the space will converge to the average initial 
temperature.

\subsection{Dispersing the Difference}
Let the manifold $\M$ be a closed square subset of $\R^2$.  
Then, we can think of the functions $f,g  \colon \M \to \R$ as surfaces 
in $\R^{3}$.  Now, let $u_0$ be the difference $g-f$.  If $u_0(x)=0$ for all 
$x \in M$, then we have $f=g$.  Otherwise, define the \emph{average} of a function as
the integral divided by the area.
$$ \text{avg}(f) = \frac{\int_M f(x) dx}{\text{area}(\M)} $$ and 
$$\text{avg}(g) = \frac{\int_M g(x) dx }{\text{area}(\M)}.$$
Then, we can calculate the average value of $u_0$ over
the domain $\M$ by subtraction: 
$$ \text{avg}(u_0)= \text{avg}(g)-\text{avg}(f).$$  We
apply the heat equation to $u_0$ and obtain $u(x,t)$ where $u(x,0)=u_0(x)$ and
$\lim_{t\to \infty}{u(x,t)}=c$.
For sake of simplicity, we will assume that the average of $u_0$ vanishes.  If
this is not the case, we can impose this condition by setting 
~$g=g-\text{avg}(u_0)$.

Now, we can observe the difference $u$ disperse through time until $u$ becomes 
the zero function.  Similarly, we know that $f(x)+u(x,t)$ will go from $g$ to 
$f$.  
Although the value of $u(x,t)$ approaches zero for all $x$ as $t$ increases
$$\lim_{t\to \infty}{u(x,t)}=0,$$ we will stop at a time $T$ when 
$u(x,T)\in (-\epsilon,+\epsilon)$ for all $x \in \M$ and for some $\epsilon > 0$.  
Then, the function 
$f+u$ goes from $g$ to a function close to $f$.  Furthermore, the 
manner in which the function $f+u$ changes is dictated by the 
heat equation.

\subsection{The Continuous Heat Equation}\label{ss:contHeat}
Here, we describe the heat equation as it applies to a continuous function.  
For more details, please refer to ~\cite{alt, numAnaly}.  In 
\secref{ss:discreteHeHo}, we will modify these equations to approximate the 
solution in the discrete case and we introduce the heat equation homotopy.

Let $\{b_1, b_2\}$ be an orthonormal basis for $\M$.
The general form of the heat equation satisfies the following conditions:
\begin{equation}\label{firstPDE} 
  \frac{\partial u}{\partial t}(x,t) - \frac{\partial^2 u}{\partial b_1^2}(x,t) - \frac{\partial^2 u}{\partial b_2^2}(x,t) = 0
\end{equation}
and the initial condition: 
\begin{equation}\label{lastPDE} u(x,0)=g(x)-f(x), x \in \M.\end{equation}
If we can solve this partial differential equation (PDE), we 
obtain $u(x,t)$ defined for all $x \in \M$ and $t \geq 0.$  Typically, the heat
equation has the additional constraint that $u(x,t)$ is constant with respect to $t$
for all $x \in \partial \M$.  However, we are interested in the case where $u(x,t)$
is \emph{heat conserving}.  That is, we would like avg$(u_{t_1})=$avg$(u_{t_2})$
for all $t_1$, $t_2$.  As we will show, in order to obtain this goal, the 
values on the boundary will reflect the values interior to the boundary.

The equation in
\eqref{firstPDE} describes how the temperature changes with respect to
time and space.
We note that if we 
impose the condition $\frac{\partial u}{\partial t}(x,t)=0$, then the heat
equation will not change with respect to time, and Equation 
~(\ref{firstPDE}) becomes Laplace's equation, $\Delta u(x,t) = 0$.  
This is known as the 
steady-state heat equation and will have a unique solution.  The iterative methods
that we look at in \secref{ss:discreteHeHo} aim at finding an approximation
of $u(x,t)$ for this problem.  The final solution will be constant with respect to
time, and so we say it is approaching the steady-state.  We are interested
in following the behavior of heat equation as it approaches the solution to the
steady-state heat equation.

\subsection{The Discrete Heat Equation Homotopy}\label{ss:discreteHeHo}
Solving a partial differential equation is not a simple task.  Thus, we must 
defer to numerical methods to estimate this solution, which require spatial and
temporal discretization ~\cite{numAnaly}.  In the following computations, we
use a regular grid decomposition of $\M = [0,1]^2$, writing $x_i=(i-1)h$ and
$y_j=(j-1)h$, where $h=1 / (n-1)$ for some fixed integer $n \geq 2$.  In the 
next section, we explain how to apply the heat equation on different topologies. 

The first step in creating the heat equation homotopy $f_t(x)$
is to compute $u_t(x) = u(x,t)$ over the discretized domain.
There are three issues that can arise when using the continuous formulation of 
the heat equation described in \secref{ss:contHeat}.
\begin{enumerate}
\item[1.] We need to solve the partial differential equation presented 
          in Equations (\ref{firstPDE}) and (\ref{lastPDE}).
\item[2.] The partial derivative $\frac{\partial^2 u_t}{\partial b_i^2}(x)$ for 
$b_i$ is not well defined over a discrete domain.
\item[3.] The solution $u(x,t)$ is defined for all non-negative $t,$ but a
          homotopy must to be defined for $t\in [0,1].$   
\end{enumerate}
In order to resolve these issues, we apply 
temporal and spatial discretization as well as scaling.  The goal is to obtain a
homotopy $f_t(x)$ from $g$ to $f$ using the heat equation solution $u(x,t).$  
Below, we describe 
one such resolution; note, however, that other approaches may be taken.

\subsubsection{Mathematical Description of the Heat Equation}
\begin{figure}[hbt]
 \vspace*{0.1in}
 \centering
 \centerline{\epsfig{figure=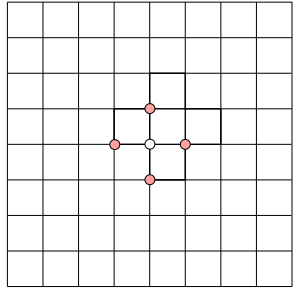,height=1.5in}}
 \caption{In the center, we have the white dot representing the mesh at position
 $x=(i,j)$.  The mesh points highlighted in pink are those whose values contribute
 to the estimate of the heat equation at $x$.}
 \label{fig:neighbors}
\end{figure}

Let us recall the steady-state heat equation over $\R^2$:
\begin{equation}\label{eq:discHeq}
 \frac{\partial^2 u}{\partial b_1^2}(x,t) + \frac{\partial^2 u}{\partial b_2^2}(x,t) = 0.
\end{equation}
In this equation, we are using $\{b_1,b_2\}$ as the standard basis vectors for $\R^2$.
At each mesh point $x=(i,j)$, we employ the Taylor polynomial in the 
variable $b_1$ to obtain an approximation of the second derivative with respect
to $b_1$:
\begin{equation}\label{eq:deriv_b1}
\frac{\partial^2 u}{\partial b_1^2} u((i, j), t)= 
\frac{u(({i+1}, j), t) -2u((i, j), t) + u(({i-1}, j), t)}{h^2},
\end{equation}
where $h$ is the spatial step size.
Similarly, we also have an approximation for the second derivative with respect
to $b_2$:
\begin{equation}\label{eq:deriv_b2}
\frac{\partial^2 u}{\partial b_2^2} u((i, j), t)= 
\frac{u((i, {j+1}), t) -2u((i, j), t) + u((i, {j-1}), t)}{h^2}.
\end{equation}
To simplify notation, we will now use $x_{i,j}$ to denote $(i, j)$.
If we plug \eqref{eq:deriv_b1} and \eqref{eq:deriv_b2} into \eqref{eq:discHeq},
then we obtain the following equation:
\begin{equation}\label{eq:linearEquations}
4u(x_{i,j}, t) - u(x_{i+1, j}, t) - u(x_{i-1,j}, t) - u(x_{i,j+1}, t) - u(x_{i,j-1}, t)
  = 0
\end{equation}
Thus, the approximation to the heat equation is made by looking at local
neighborhoods for each point $x_{i,j}$.
\figref{fig:neighbors} highlights the four neighbors of the mesh needed to
compute an approximation to the heat equation.
The neighborhood of a point, Nhd$(i,j)$, is defined to be the set of local
neighbors of the point $x_{i,j}$.
\begin{equation}\label{eq:nhd}
 \text{Nhd}(i,j) = \{(i,j \pm 1), (i \pm 1,j) \} 
\end{equation}
We have $n^2$ equations of the form presented in \eqref{eq:linearEquations}, 
one for each point in the $n \times n$ mesh.
We relabel the mesh points in column-major order in order to use one 
index instead of two:
$ v_{\{(j-1)n + i\}} := u(x_{i,j}, t)$.
Then, we may express the $n^2$ 
linear equations in matrix-vector form, ~$Av=0$.

\subsubsection{An Iterative Algorithm for Linear Systems of Equations}\label{ss:jacobi}
As we have shown above, solving the discrete heat equation finds a solution $v$
to the linear system of equations $Av=0$.
Above, we have described how to construct the matrix $A$, as it is the
coefficient matrix for the system of linear equations.  In the above description,
$A$ is equal to $L_n$, the \emph{Poisson matrix} of order $n$.  We note here 
that this matrix is $n^2 \times n^2$.
We can write $L_n = D-N$, where $D$ is the diagonal matrix $4\cdot I$ 
and $N$ is a matrix with $0$'s on
the diagonal and with only $1$ as the non-zero entries of the matrix.  Sometimes
we refer to $D$ as the \emph{valency matrix}, since it expresses the degree of
each mesh point. The matrix
$N$ is symmetric and we call it the \emph{neighborhood matrix} since the 
non-zero entries in row $i$ correspond to the neighbors of the mesh
point $v_i$ ~\cite{babicResDist}.  That is, $N(i,j)=1$ iff $v_i$ and $v_j$ 
are adjacent.

The iterative algorithm can be defined by these matrices.  We want a solution 
of the form $Av=0$, which means $(D-N)v=0$.  We can re-write this so that $Dv=Nv$.
And, the iterative algorithm can immediately be seen:
\begin{equation}
 v_{\text{new}} = (D^{-1}N)v.
\end{equation}
In the original formulation, this translates to:
$$ u_{t+1}(x) = \frac{1}{4} \sum_{y \in \text{Nhd}(x) } u_t(x),$$
where the neighborhood Nhd$(x)$ is the neighborhood of $x$ defined by 
Equation \eqref{eq:nhd}.  This iterative method is known as 
\emph{Jacobi iteration}.

\subsubsection{Creating the Homotopy}
We continue the process of obtaining $u_{t+1}$ from $u_t$ until we have 
reached a halting point, where $|u_{t+1}(x)-u_{t}(x)| \leq \epsilon$ for all
$x \in \M$ and for some predetermined value of $\epsilon$.  We will let $T$ be the
maximum time computed in the iterative method.
Thus, we have a function ~$u:\M \times [0,T] \to \R.$  
We reparameterize 
$u(x,t)$ with respect to $t$ in order to obtain $\tilde{u}(x,t)=u(x,{t}\cdot {T})$ 
defined over the domain $\M \times [0,1].$  Then, the heat equation 
homotopy $f_t$ can be defined by the equation: 
\begin{equation}\label{eq:heho}f_t(x)=f(x) + \tilde{u}(x,t).\end{equation}
Notice that $f_0(x)=g(x)$ and 
$f_1(x) \approx f(x)$ with $T$ sufficiently large since 
$$\lim_{T \to \infty} f_1(x)= \lim_{T \to \infty} f(x) + u(x, T)= f(x).$$  
By using this homotopy,
the initial difference between $f$ and $g$ disperses.  Although $\tilde{u}(x,t)$
approaches a constant function as $t$ approaches $T$, interesting things can
happen along the way.  For example, critical values can be created.

\begin{figure}[hbt]
 \vspace*{0.1in}
 \centering
 \centerline{\epsfig{figure=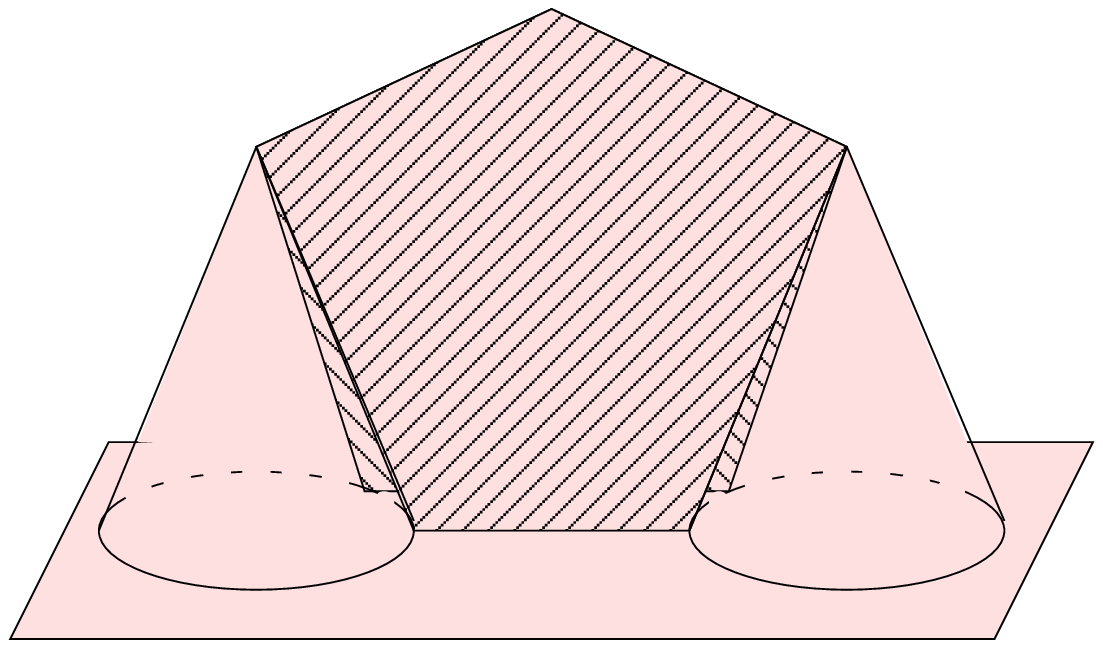,height=1.5in}}
 \caption{The function used in \exref{ex:mountain}.}
 \label{fig:mountain}
\end{figure}
\example[Mountain and Bridge]\label{ex:mountain}
Suppose we have two cones connected by two pentagons as shown in 
\figref{fig:mountain}.  We will call this surface $S$.
Now, consider the height $h \colon \R^2 \to \R$ defined to be the height
of the surface at $(x,y)$ below $S$ and zero elsewhere.  We assume that the 
base of the cones and the pentagons is in the $xy$-plane.
There is one dimension one critical value, the pinnacle of the pentagons.
If we apply the heat equation to this surface, there will soon be at least
three dimension zero critical values: one corresponding to the original 
critical value and two from the cone tips.

\subsection{Considering Different Topologies}
Here, we consider implementing different topologies
for the square domain.
Each mesh element has four neighbors, corresponding to above,
below, left, and right.
The topologies are determined by which vertices we define to be neighbors.
We consider four topologies: the square, the torus, the Klein bottle, and the 
sphere.  
The square topology, used when describing the iterative
heat equation, is the most basic.
\begin{definition} [Square Topology]
  The neighborhood of $(i,j)$ is: 
$$ Nhd(i,j) = \{ (i, j \pm~1),   (i \pm 1, j) \} ,$$
provided that these elements are within the domain.
\end{definition}
Each vertex can have 
two, three, or four neighbors, resulting in a system of equations that does
not conserve heat.
Let $V$ be the set of vertices in $\M$, and define the total heat to be the
sum of the heat over the entire domain:
~$\sum_{x \in V}u(x).$
Then, the total heat is not preserved between iterations.  To fix this, let 
$x$ be its own neighbor for every neighbor
that $x$ is missing.  Hence, each $x \in V$ is the neighbor of four other
vertices and has four neighbors itself.  For the second step, we have
$$\sum_{x \in V}u(x,t+1) =\sum_{x\in V}\frac{4u(x,t)}{4} 
                        =\sum_{x\in V}u(x,t).$$

\begin{figure}[ht]
	\vspace*{0.1in}
	\centering
	\centerline{\epsfig{figure=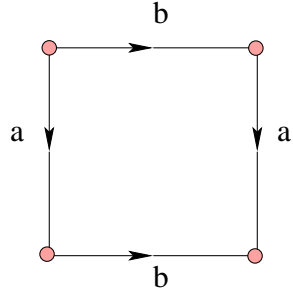,height=1.5in}}
	\caption{The torus is the quotient of the unit square by gluing together
        opposite sides as prescribed by their orientation.}
	\label{fig:torus}
\end{figure}

The four edges of the square create the boundary on the square topology.
If we identify the boundary edges in pairs, we can then create a surface 
without boundary.  One way to do this is to create a torus from the square.
Formally, we define the torus topology as follows:
\begin{definition} [Torus Topology] 
  Given an $n \times n$ mesh, the neighborhood of $(i,j)$ is: 
  $$Nhd(i,j) = \{(i, j \pm_n 1), (i \pm_n 1, j)\},$$
 where addition and subtraction are calculated modulo $n$.
\end{definition}
The torus topology can also be described by \figref{fig:torus}. 
\begin{figure}[ht]
	\centering
	\subfigure[The Klein bottle]{
	\includegraphics[scale=1]{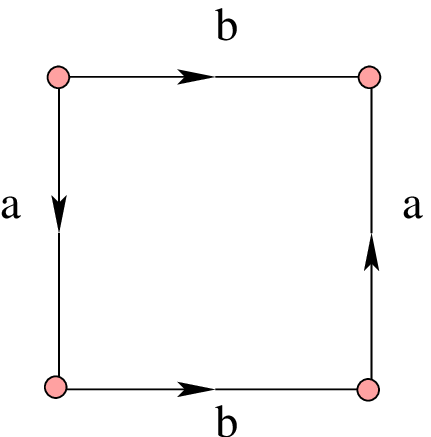}
	\label{fig:klein}
	}
	\subfigure[The sphere]{
	\includegraphics[scale=1]{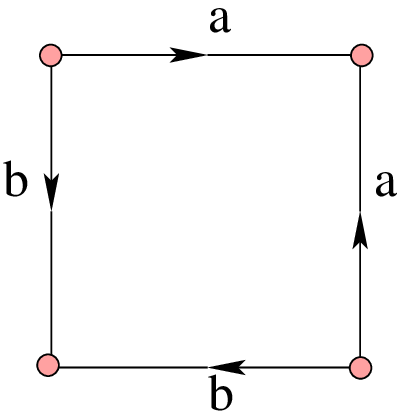}
	\label{fig:sphere}
	}
	\caption[Klein Bottle and Sphere Topologies]{The figures above
	illustrate how the boundaries of the square are glued to obtain
	the Klein bottle and the sphere.}\label{fig:kbNsphere}
\end{figure}
By gluing the two $a$ edges together and the two $b$ edges together, we obtain
a manifold without boundary.  In this manifold, each vertex $v$ is the neighbor
of four other vertices, and has four neighbors that contribute to the new value
for that vertex.
Similarly, we can define the Klein Bottle and the Spherical topologies
as depicted in \figref{fig:kbNsphere}.

\subsection{Convergence}
When using iterative algorithms, we worry about how
long finding the solution (or something close to the solution) will take.  In 
this section, we discuss one metric that measures the convergence of iterative
algorithms.

If $x^{(k)}=u(x,t_k)$ is the solution
at stage $k$, then the solution at stage $k+1$ is 
$$ x^{k+1} = \frac{1}{4}Nx^{(k)}.$$
We now see that
\begin{equation}\label{eq:iter} Dx^{(k+1)}=Nx^{(k)}\end{equation}
and for the solution $x^{\infty}$:
\begin{equation}\label{eq:solution}Dx^{\infty}=Nx^{\infty}.\end{equation}
If we subtract (\ref{eq:iter}) from (\ref{eq:solution}), then we obtain
\begin{equation*} DE_{k+1} =NE_k,\end{equation*}
where $E_k=x^{\infty}-x^{k}$ is the error of the $k^{th}$ estimate $x^{(k)}$.
Hence, $$E_{k+1}=D^{-1}NE_k=(D^{-1}N)^{k+1}E_0.$$  The action 
of $M=D^{-1}N$ on the initial error determines whether the error of the 
solution $x$ will increase or decrease.  Now, if $M$ has $n^2$ non-degenerate
eigenvalues and $n^2$ linearly independent eigenvectors, we can write
$M=V \Sigma V^{-1}$, where $\Sigma$ is the diagonal matrix of eigenvalues.
Since $M^k = V \Sigma^k V^{-1}$, we use the 
spectral norm of the matrix $M$ to determine if the error is compounding or 
decreasing.  The \emph{spectral norm}, $\rho$, of a matrix the largest absolute
eigenvalue of that matrix.  If $\rho<1$, then the iterative algorithm 
converges.  If $\rho>1$, then the iterative algorithm does not converge.
For example, if we are using Jacobi iteration as described in 
\secref{ss:jacobi}, then $\rho$ is very close to 
one, and thus convergence is very slow ~\cite{compSim}.
Although this is undesirable behavior for finding a solution to the heat
equation, for our purposes, it allows us to more closely examine the behavior
of the heat equation.

\section{Vineyards of the Heat Equation Homotopy}\label{sec:hehovin}
In \secref{ss:vineyard}, we defined homotopy and stated that we can stack the
persistence diagrams associated with the homotopy to create a vineyard.
In this section, we estimate the underlying vineyard for the heat equation
homotopy by monitoring the transpositions in the filter.

\subsection{Turning a Mesh into a Filter}
We begin with an $n \times n$ matrix of values.  Although the heat equation
computes neighbors based on a grid, we will be computing persistence using
simplicial complexes.  Thus, we triangulate the domain
by adding in horizontal, vertical and diagonal edges, as well as the
triangles formed by the voids.  

\begin{figure}[hbt]
 \vspace*{0.1in}
 \centering
 \centerline{\epsfig{figure=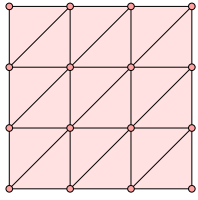,height=1in}}
 \caption{The triangulation of a $4 \times 4$ grid.}
 \label{fig:mesh}
\end{figure}
An $n \times n$ mesh gives us $n^2$ vertices, $(n-1)(3n-1)$ edges, 
and $2(n-1)^2$ faces.  In \figref{fig:mesh}, we see the triangulation
of a four-by-four mesh.  Each simplex $\sigma$ is defined by the vertices that
create it.  In addition, the value of a simplex, $f(\sigma)$, is the maximum 
function value of those vertices.  Now, we order the simplices by the following 
two rules:
\begin{enumerate}
   \item[1.] If $f(\sigma_1) < f(\sigma_2)$.
         then $\sigma_1$ appears before simplex $\sigma_2$. 
   \item[2.] If $\tau$ is a subsimplex of $\sigma$, denoted $\tau \leq \sigma$, 
         then simplex $\tau$ appears before simplex $\sigma$.
\end{enumerate}
We observe here that the second rule does not contradict the first, because
$f(\tau) \leq f(\sigma)$ whenever $\tau$ is a face of $\sigma$.  The resulting
ordering of the simplices is called a \emph{filter}.  The two rules imply that
every initial subsequence of the filter defines a subcomplex of the mesh.
Growing this initial subsequence until it equals the entire filter gives
a sequence of simplicial complexes called the induced \emph{filtration}.
As with any sorting algorithm, to create a filter
of $m$ simplices will take $O(m\log m)$ time.

\subsection{Computing Vines}\label{ss:vines}
Recall the heat equation homotopy from Equation \eqref{eq:heho},
$$f_t=f(x) + \tilde{u}(x,t),$$ which is defined over
a finite number of time-steps: $0, t_1, t_2, \ldots, t_T=1$.
Create a filtration for $f_0(x)$ as prescribed above.
We use the filtration to compute the persistence diagrams, just as we used the 
sublevel sets in \secref{ss:persHomology}.
We progress from one complex in the filtration to another by adding one simplex.
The addition of this simplex can be the birth of a new homology class or the
death of an existing homology class.  After we have iterated through the
entire filter, we have
completed the computation of the persistence diagrams for time ~$0$.

To compute Dgm$_p(f_{t_1})$, we could repeat the same process.  However, if we
use Dgm$_p(f_{t_0})$, we can compute the new diagram in linear time and we can
match points in the two diagrams ~\cite{vines}.

\subsubsection{Transpositions}\label{ss:transpose}
Suppose we have two filters on $m$ simplices, such that the filters are identical 
except that two adjacent simplices have swapped order.  
Then, we can write the first filter:
$$ F1: \sigma_1, \sigma_2, \ldots, \sigma_i, \sigma_{i+1}, \ldots, \sigma_m$$
and the second filter:
$$ F2: \sigma_1, \sigma_2, \ldots, \sigma_{i+1}, \sigma_{i}, \ldots, \sigma_m.$$
If we have the persistence diagram
for the filter $F1$, then we can compute the persistence diagram for the
the filter $F2$ by performing one \emph{transposition}.  A transposition 
updates the persistence diagram by changing the pairings of two consecutive 
simplices, if necessary.  
In the case that $\text{dim}(\sigma_{i}) \neq \text{dim}(\sigma_{i+1})$,
the transposition does not affect the pairs in the persistence diagram.  
Thus, only the transposition of simplices with the same number of vertices
can result in a pairing swap. One transposition can swap the births and deaths
of at most two points in the persistence diagram.

\begin{figure}[hbt]
 \vspace*{0.1in}
 \centering
 \centerline{\epsfig{figure=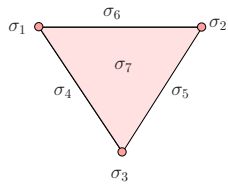,height=1in}}
 \caption{The filter of a simplicial complex is an ordering on the vertices,
           edges, and faces.
	  The simplices are originally ordered $\sigma_1$ through ~$\sigma_7$.
          Swapping $\sigma_1$ and $\sigma_2$ results in a pairing swap of type ~$1$.
          Swapping $\sigma_4$ and $\sigma_5$ results in a pairing swap of type ~$2$.
          Swapping $\sigma_5$ and $\sigma_6$ results in a pairing swap of type ~$3$.}
 \label{fig:transpose}
\end{figure}
We must distinguish between nested and unnested persistence pairings.
We say that two persistence points, $(b_1, d_1)$ and $(b_2, d_2)$, are 
\emph{nested} if the birth at $b_2$ and the death at
$d_2$ occur after $b_1$ and before $d_2$.  Assuming all events happen at
distinct moments of time, this is equivalent to ~$b_1 < b_2 < d_2 < d_1$.
If we transpose $\sigma_i$ and $\sigma_{i+1}$, then
the three types of pair-swapping transpositions are:
\begin{enumerate}
 \item[1.] The births of two nested pairs are transposed.  In this case,
           $\sigma_i$ is associated with $b_1$  in Dgm$_p(F1)$, and with
           $b_2$ in Dgm$_p(F2)$.
 \item[2.] The deaths of two nested pairs are transposed.  In this case,
           $\sigma_i$ is associated with $d_2$  in Dgm$_p(F1)$, and with
           $d_1$ in Dgm$_p(F2)$.
 \item[3.] The birth of one pair and the death of another, unnested, pair
           are transposed.  In this case, the addition of $\sigma_i$ to
           the filtration created a death in Dgm$_p(F1)$, but a birth
           in Dgm$_p(F2)$.
\end{enumerate}
\figref{fig:transpose} illustrates the three types of transpositions 
that can be made.  Although pair swaps of types $1$ and $2$ involve
two persistence points in the same diagram, pair swaps of type $3$
involve persistence points in diagrams of two consecutive dimensions.
In each of the cases, the transposition results in a swap only if
changing the order in which the simplices are added changes the persistence
pairing that is made.  For a complete algorithm to compute the
transpositions, please refer to ~\cite{vines}.

From these transpositions, we can create the matching referred to in
Stability Result of the Straight Line Homotopy of \ref{ss:stability}.
Suppose the transposition of $\sigma_i$ and of $\sigma_{i+1}$ resulted
in a pairing swap.  Then, every persistence point is paired with
itself in the matching, except for $(b_1,d_1)$ and ~$(b_2,d_2)$, whose pairings
are swapped.

\subsubsection{Sweep Algorithm}
Changing one filtration into another may require more than one 
transposition.  In order to compute the persistence diagram,
we first create a topological arrangement and then use a sweep
algorithm. 
\begin{figure}[hbt]
 \vspace*{0.1in}
 \centering
 \centerline{\epsfig{figure=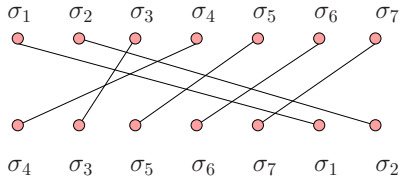,height=1in}}
 \caption{ We create a topological arrangement by connecting 
         drawing a line between the vertices that represent the same simplex.
         By doing this, we find a finite number of crossings.  Each crossing
         represents one transposition in the filter.}
 \label{fig:sweep}
\end{figure}
In the Cartesian plane, write the filtration of $f_{t_j}$ horizontally.
Below that, write the filtration of $f_{t_{j+1}}$ using the same simplex names.
Then, we connect like simplices with a single curve (does not need to be straight).
An example of this process is given in \figref{fig:sweep}.
After this arrangement has been created, an ordering on the transpositions
can be found by topologically sweeping the arrangement, as presented
in \cite{sweep}.

We compute Dgm$_p(f_{t_{j+1}})$ and keep track of the matching by progressing one 
transposition at a time in the order dictated by the sweep algorithm.

\subsection{Measuring Distance between Persistence Diagrams}\label{ss:hehoMatching}
Now, we assume that
there exists a homotopy between functions $f$ and $g.$  Then, we have a vineyard
that connects each point $a$ in $A=$Dgm$_p(f)$ with a point $b$ in 
$B=$Dgm$_p(g).$  We will use this pairing as our matching and define a distance 
metric for it. 

Assume $a$ and $b$ are connected
by the vine $s:[0,1] \to \R^3,$ as described in \secref{ss:vines}.  Since the 
vine was created from a homotopy, we will use the index $t$ to emphasize that 
$s(t)$ is a persistence point for the function $f_t(x).$
The velocity of the vine $\frac{\partial s}{\partial t}$ will be integrated on 
$[0,1]$ in order to measure the distance traveled between $a$ and $b:$
$$ D_{s} = \int_0^1  \frac{\partial s(t)}{\partial t}  \,d t.$$
In order to obtain a distance between persistence diagrams we sum these 
distances over all vines in the vineyard $V$:
\begin{equation}\label{eq:intdist} D_{fg} = \sum_{s \in V}  D_s. \end{equation}
If $s(t)$ is only defined for a discrete set of times $t_i$ with $0 \leq i \leq T$,
then we obtain an alternate definition for ~$D_{fg}$:
\begin{equation}\label{eq:discdist} 
D_{fg} = \sum_{s \in V}  \sum_{i \in (0,T]}  || s(t_i)-s(t_{i-1}) ||_{\infty}. 
\end{equation}
When a point $a \in A$ enters the diagonal at $t<1,$ we pair $a$ 
with the corresponding diagonal point, since the diagonal points can only 
occur as endpoints of a vine by definition of vine in \secref{ss:vines}.   
Then, we only measure the distance over the interval in which the vine is 
defined, $[0,t).$  Symmetrically, we can have a diagonal point in $A$
paired with an off-diagonal point in $B$.

We are interested in understanding how this distance metric compares with the 
bottleneck and Wasserstein distances, as well as to understand the properties of this distance
metric and homotopy matching.

\section{An Example}\label{sec:results}
\begin{figure}[ht] 
	\centering
	\subfigure[Bird]{
	\includegraphics[height=1in]{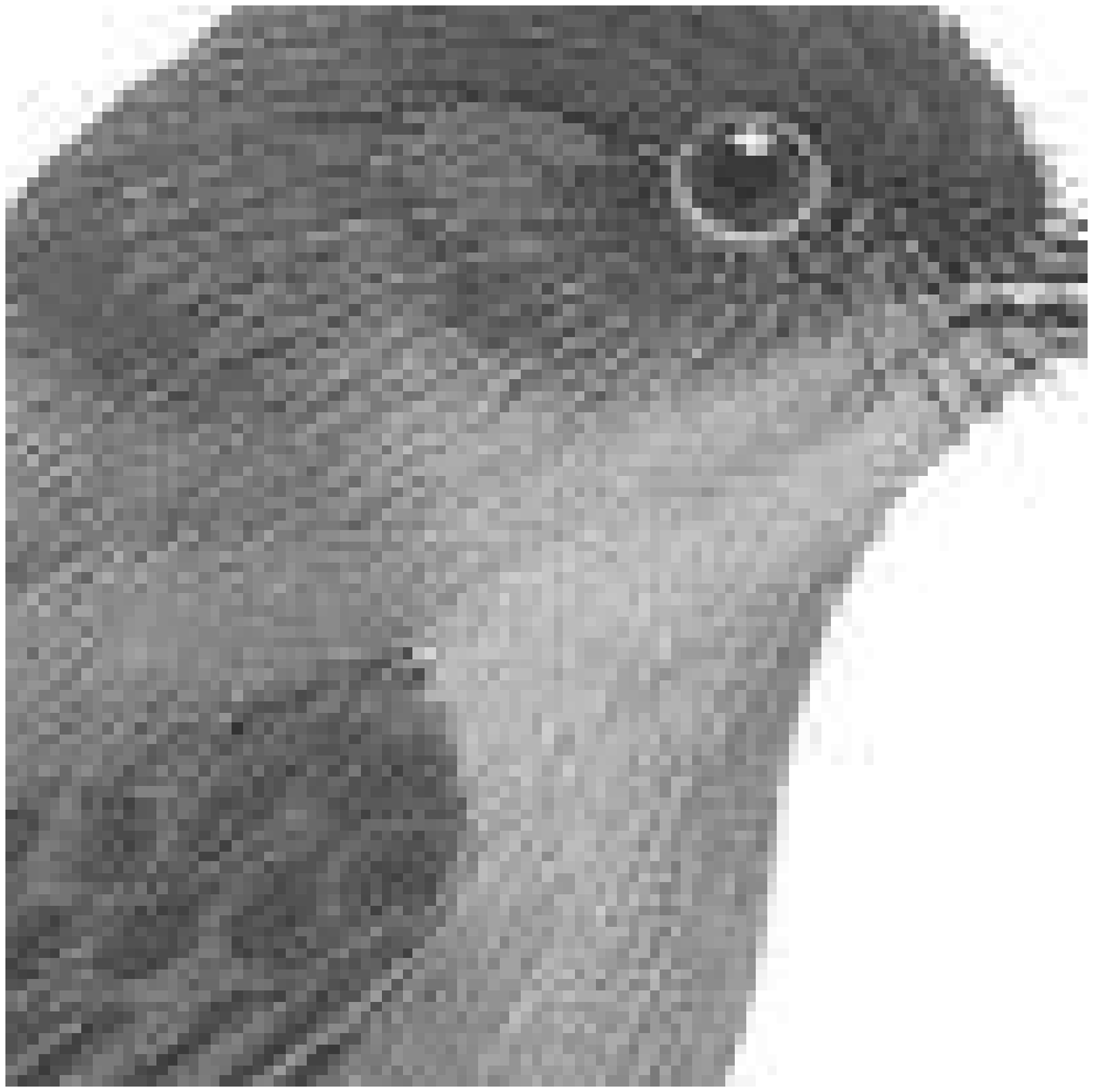}
	\label{fig:bird}
	}
	\subfigure[Flower]{
	\includegraphics[height=1in]{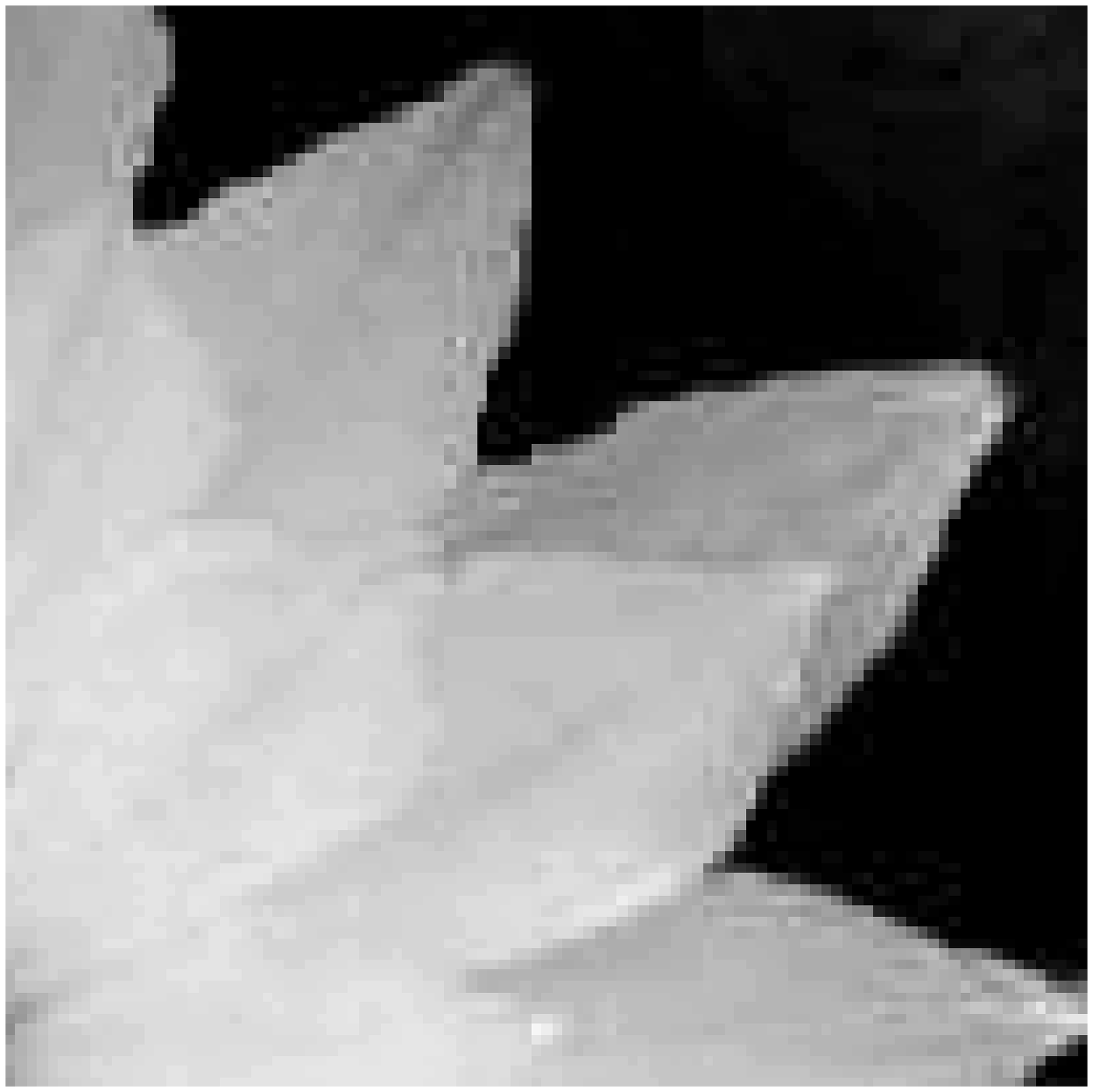}
	\label{fig:flower}
	}
	\subfigure[Difference]{
	\includegraphics[height=1in]{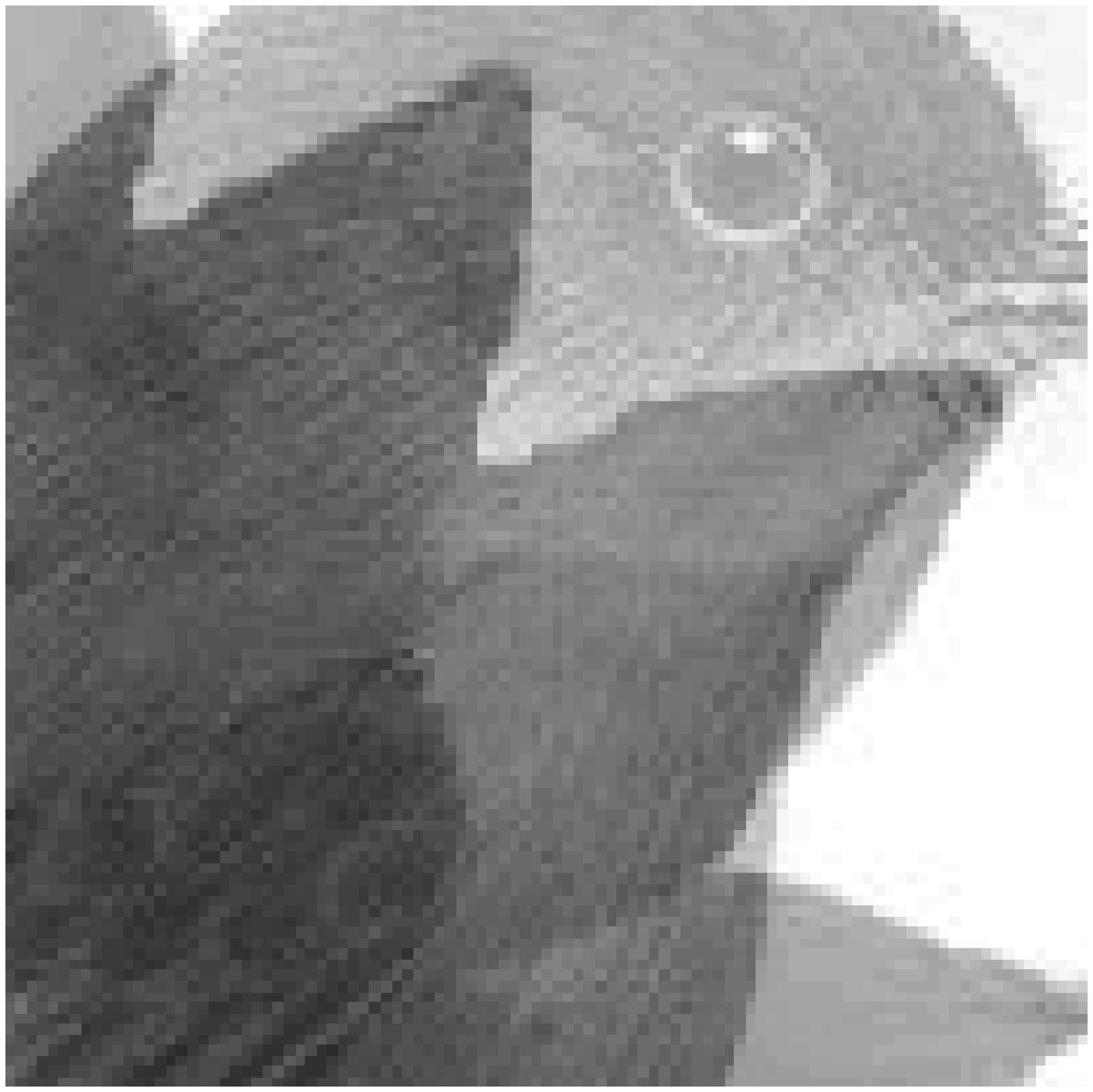}
	\label{fig:diff}
	}
	\caption[Images of the Homotopy]{The homotopy acts on the difference
                between the image of the bird and the image of the flower.
                The grayscale values of the image of the difference represent
                the values of the height function　\ $u_0$.  In the images,
                the dark pixels correspond to the low values and the light
                pixels correspond to the high values.
                }\label{fig:htpyImages}
\end{figure}

In this section, we present the results from one example.  Although the results
from one example cannot be generalized, we are able to capture the different
behaviors of the homotopy under the four topologies (square, sphere, torus, and 
Klein bottle).  In the subsequent examples, $\M$ is a square closed region of 
$\R^2.$    The functions  $f$ and $g$ are approximated by a $101 \times 101$ 
mesh.  We obtain the function values from two gray-scale images of a bird and 
a flower respectively, as shown in \figref{fig:htpyImages}.  The difference 
$u_0=g-f$ is computed.  From here, we will find a homotopy from $u_0$ to the 
zero function.  We note that this homotopy differs from the one previously
defined, since we do not add $f$ to the heat equation solution.
In practice, we found that this homotopy more clearly displays the behavior of the
heat equation.
\begin{figure}[hbt] 
 \vspace*{0.1in}
 \centering
 \centerline{\epsfig{figure=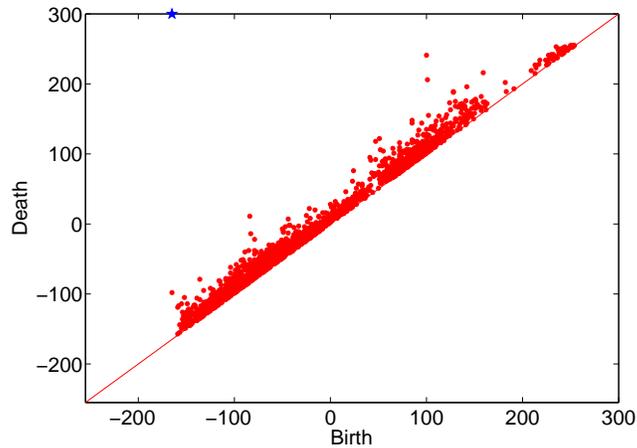,height=2.5in}}
 \caption{The persistence diagram of the difference function.  This is the diagram
          at Step $0$ of the heat equation under the square topology.  The
          possible values are the integers in ~$[-255,255]$.  The blue
          star drawn at height $300$ represents the essential homology class.}
 \label{fig-01:Sumi}
\end{figure}

\begin{figure}[ht]
	\centering
	\subfigure[Step 1]{
	\includegraphics[width=.3\textwidth]{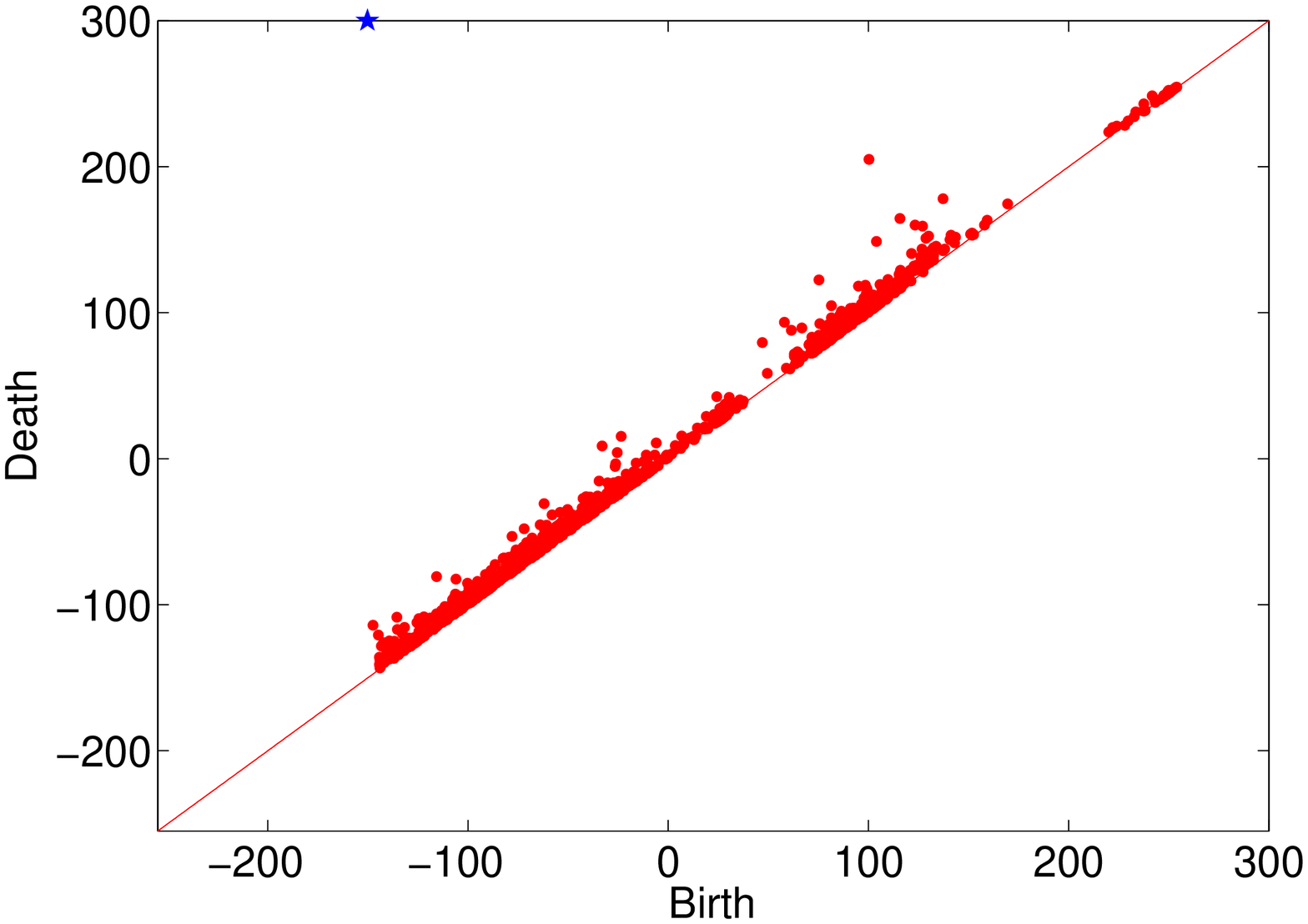}
	\label{fig:dgm-square-step1}
	}
	\subfigure[Step 10]{
	\includegraphics[width=.3\textwidth]{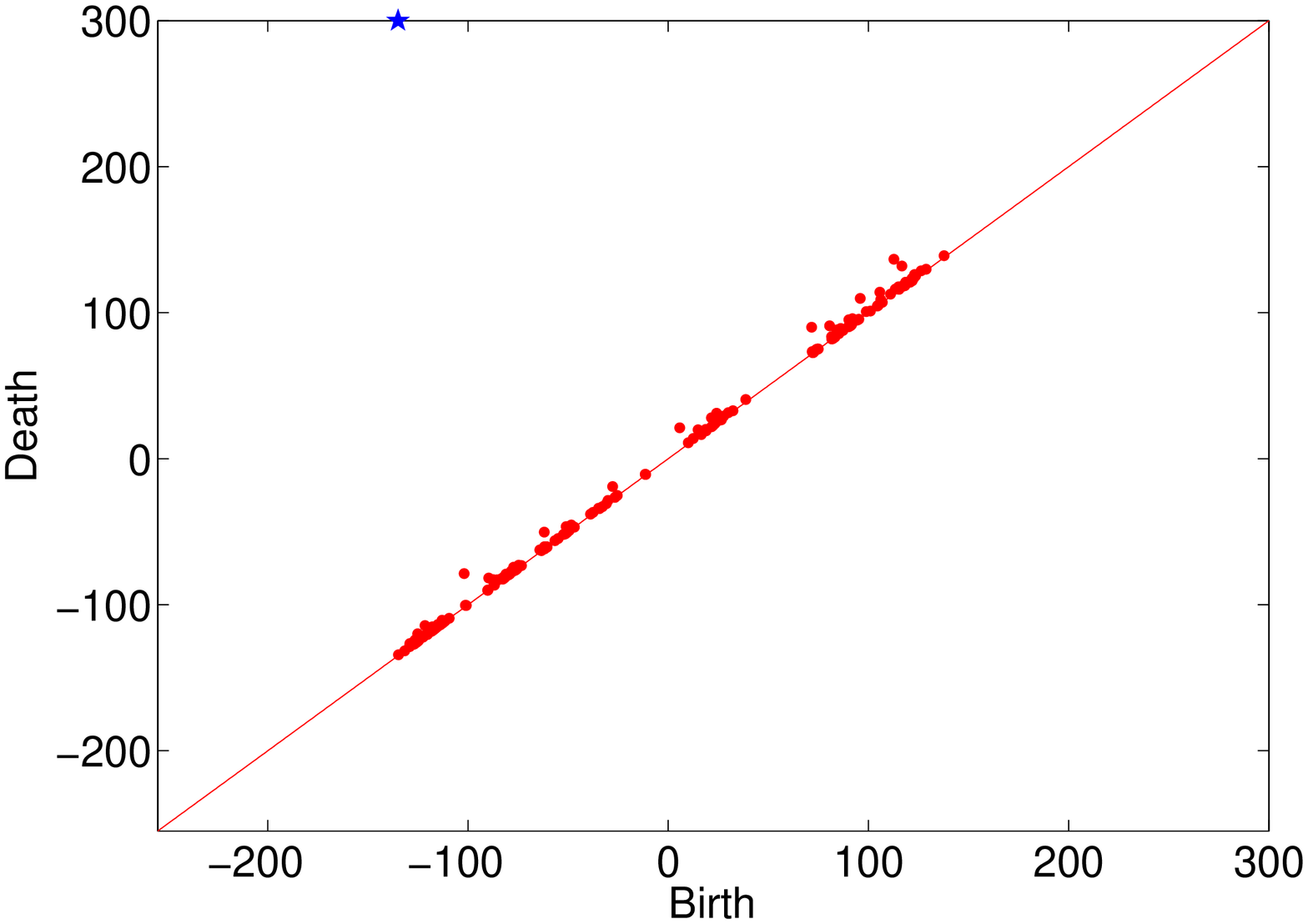}
	\label{fig:dgm-square-step10}
	}
	\subfigure[Step 100]{
	\includegraphics[width=.3\textwidth]{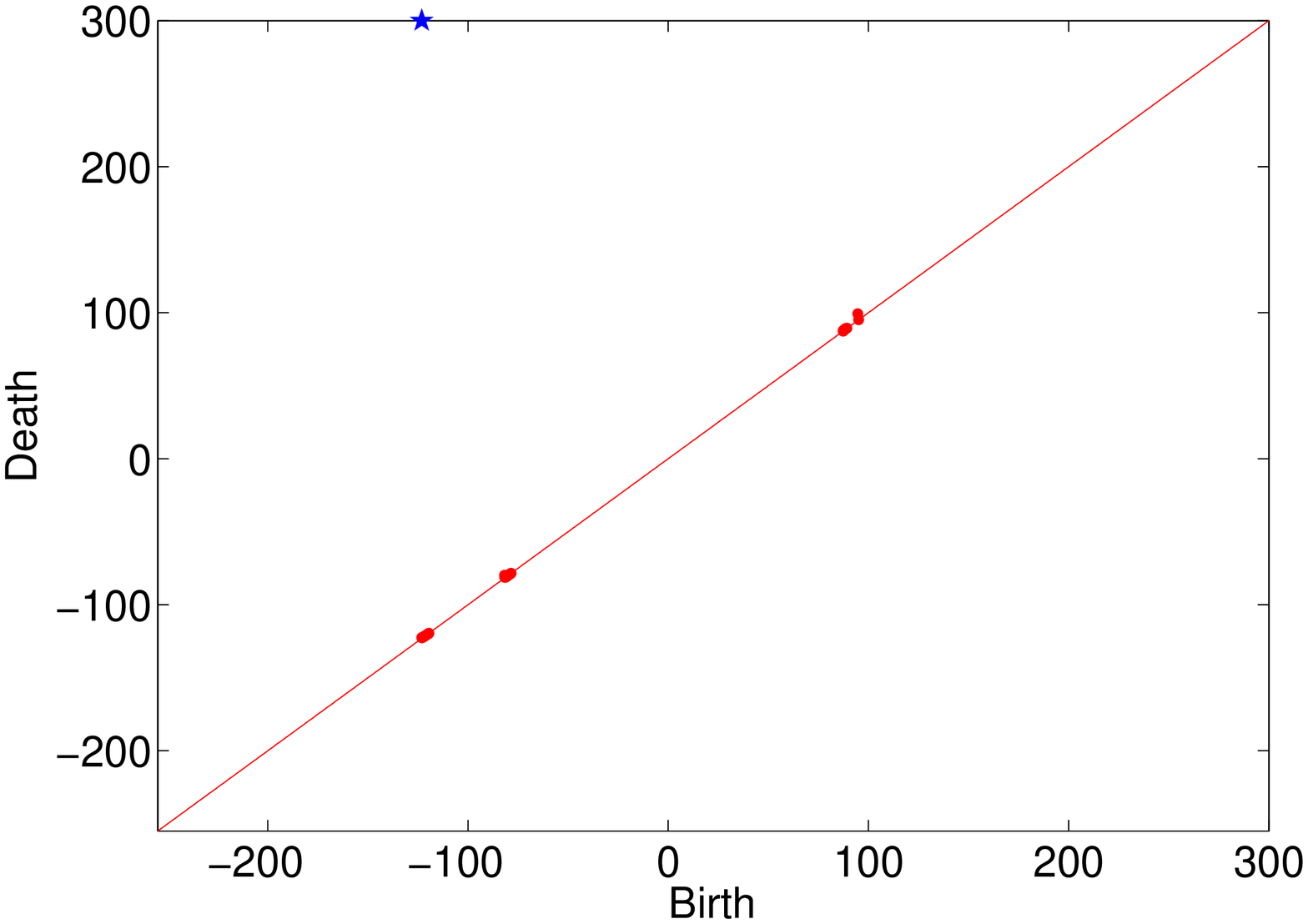}
	\label{fig:dgm-square-step100}
        }
	\caption[Torus Boundary]{Persistence diagrams for the heat equation using
                 square topology.
                }\label{fig:dgm-square}
\end{figure}
\begin{figure}[ht]
	\centering
	\subfigure[Step 1]{
	\includegraphics[width=.3\textwidth]{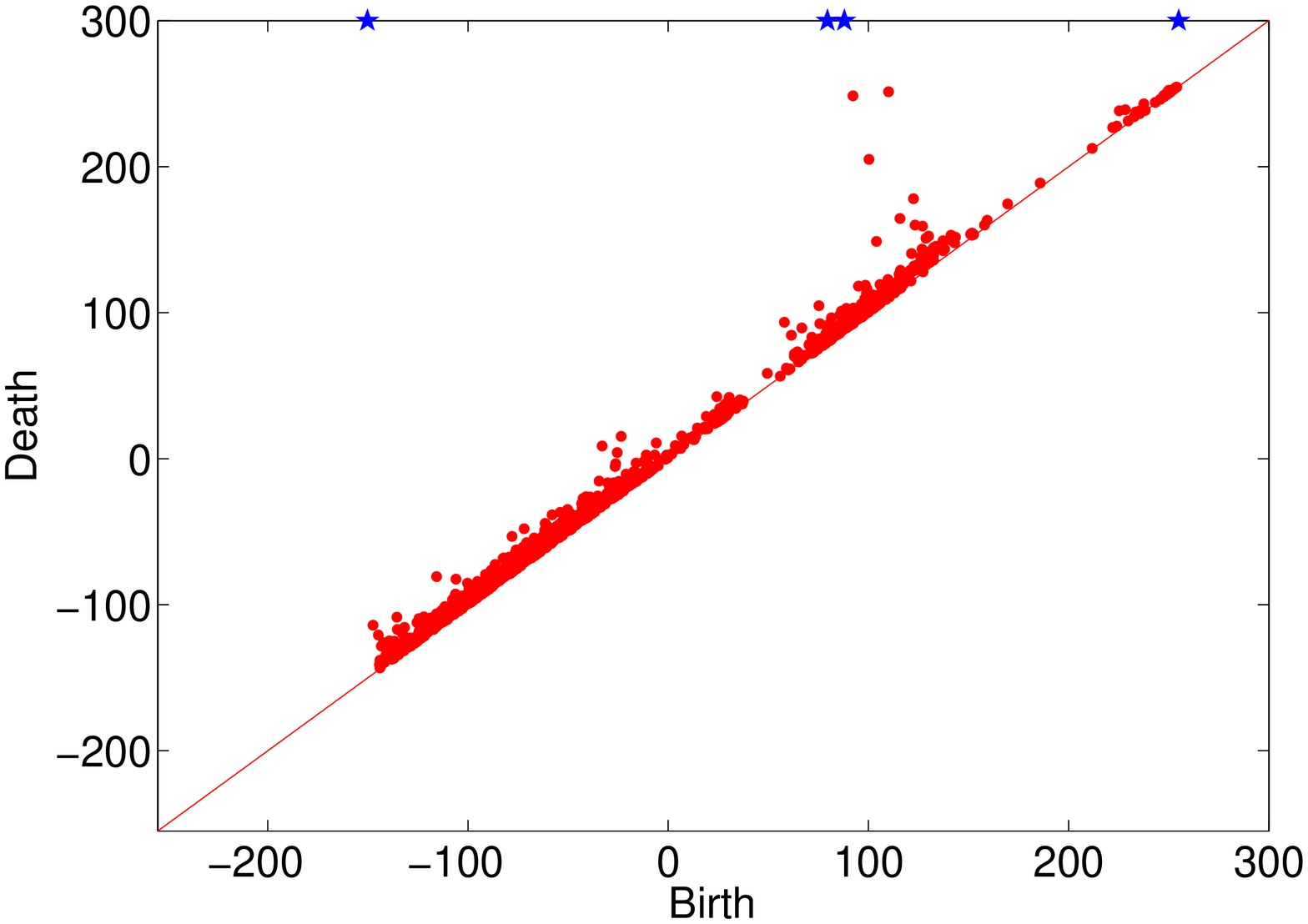}
	\label{fig:dgm-torus-step1}
	}
	\subfigure[Step 10]{
	\includegraphics[width=.3\textwidth]{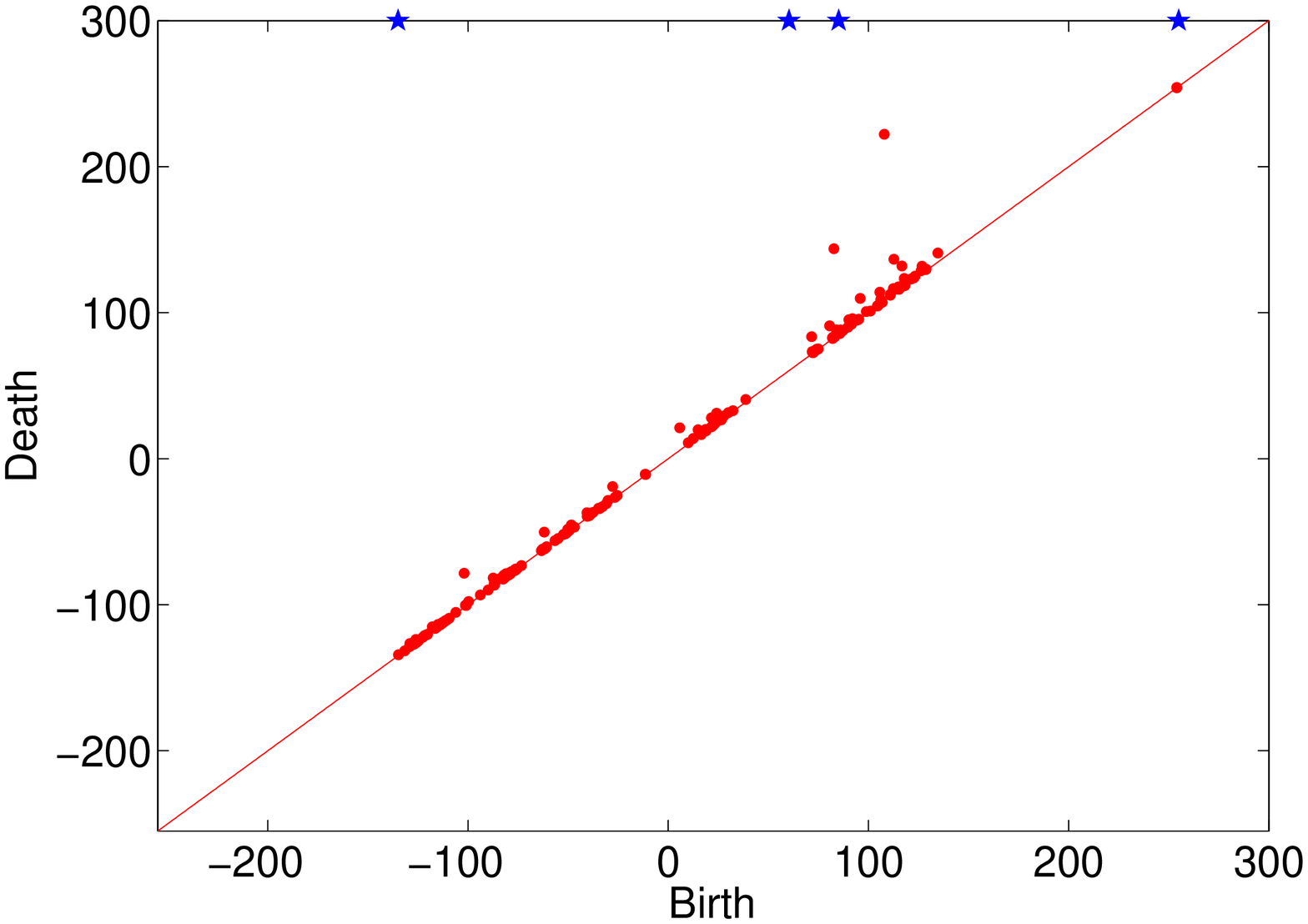}
	\label{fig:dgm-torus-step10}
	}
	\subfigure[Step 100]{
	\includegraphics[width=.3\textwidth]{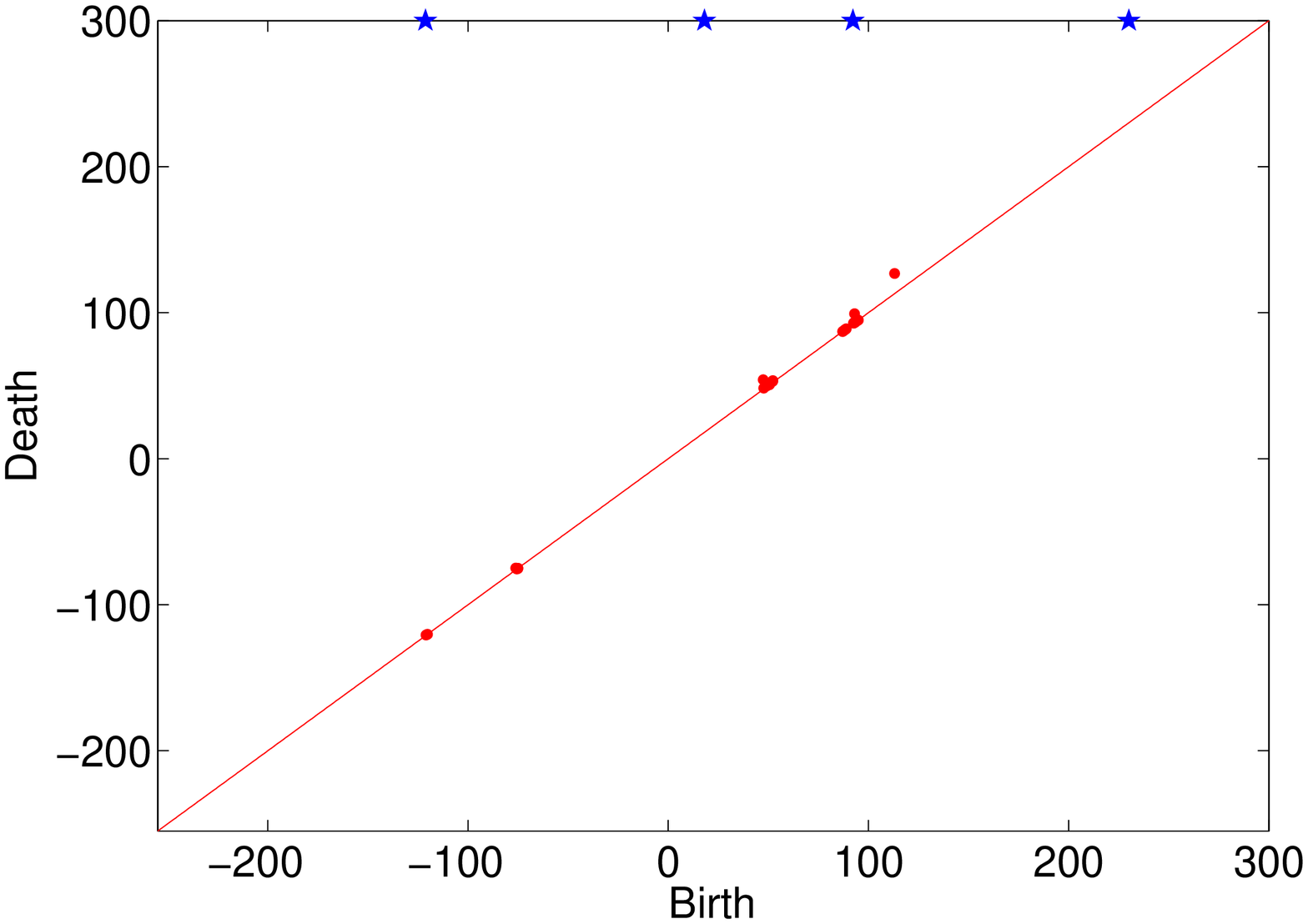}
	\label{fig:dgm-torus-step100}
        }
	\caption[Torus Boundary]{Persistence diagrams for the heat equation using
                 torus topology.
                }\label{fig:dgm-torus}
\end{figure}
\begin{figure}[ht]
	\centering
	\subfigure[Step 1]{
	\includegraphics[width=.3\textwidth]{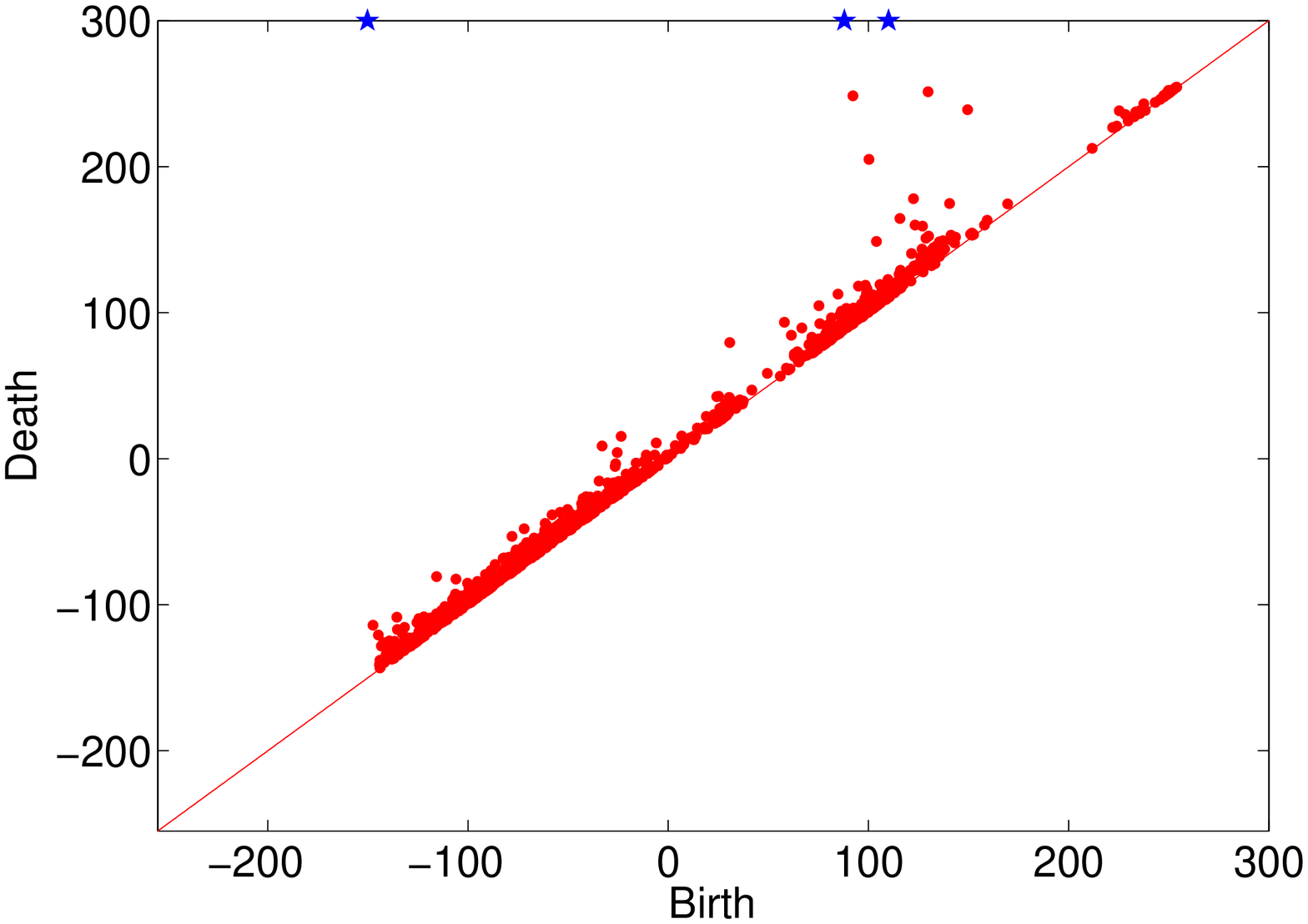}
	\label{fig:dgm-kb-step1}
	}
	\subfigure[Step 10]{
	\includegraphics[width=.3\textwidth]{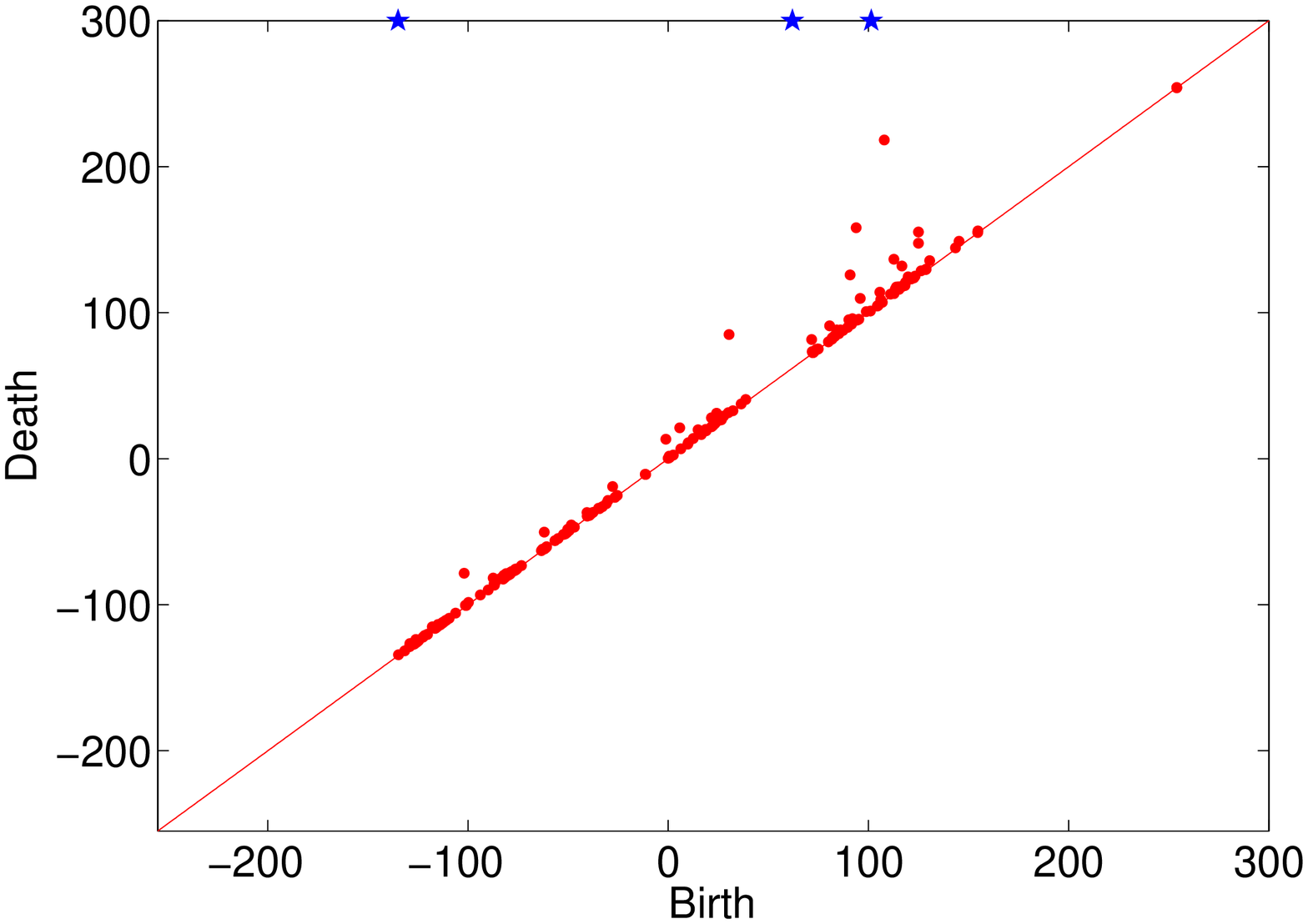}
	\label{fig:dgm-kb-step10}
	}
	\subfigure[Step 100]{
	\includegraphics[width=.3\textwidth]{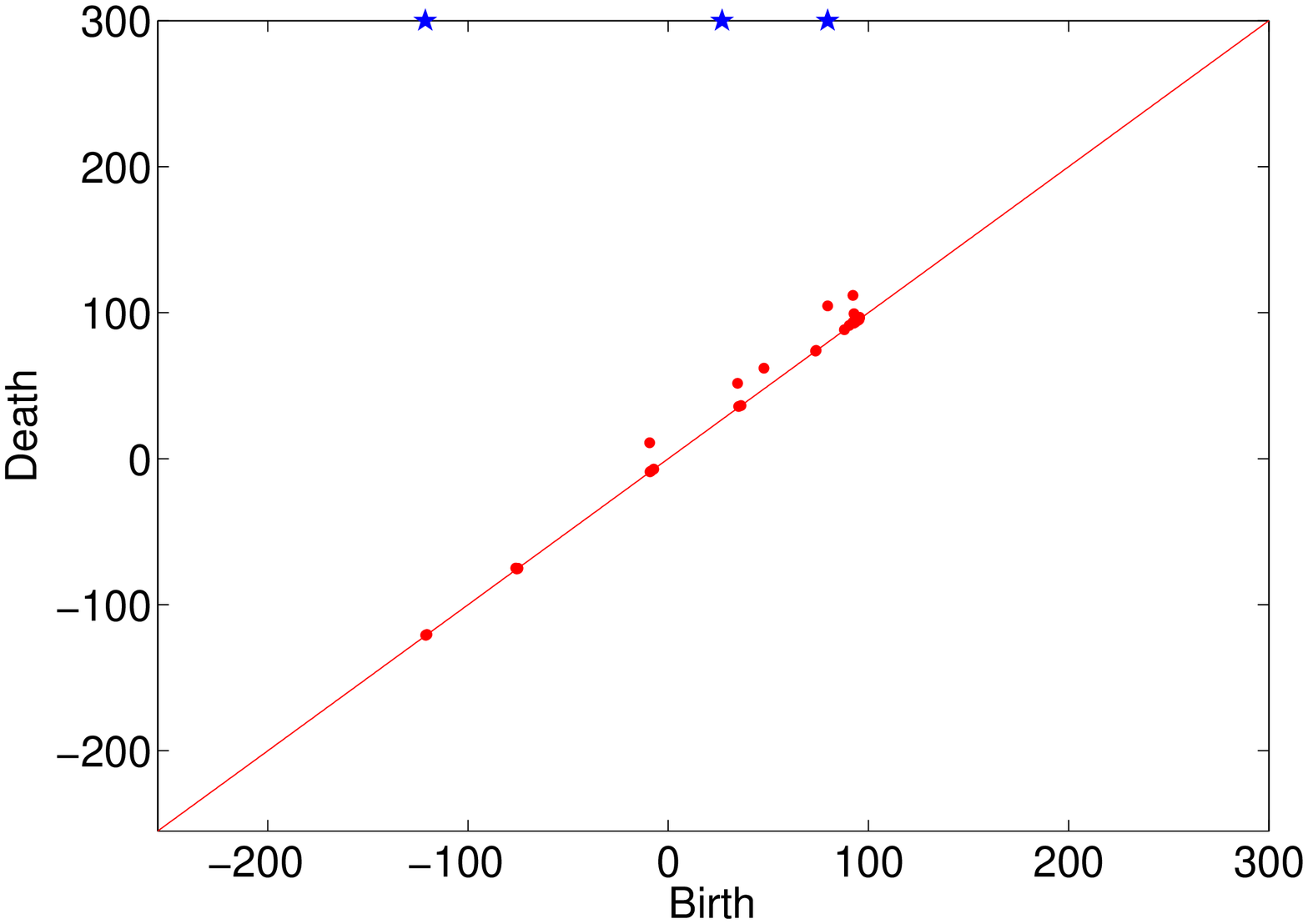}
	\label{fig:dgm-kb-step100}
        }
	\caption[KB Boundary]{Persistence diagrams for the heat equation using
                 Klein bottle topology.
                }\label{fig:dgm-kb}
\end{figure}
\begin{figure}[ht]
	\centering
	\subfigure[Step 1]{
	\includegraphics[width=.3\textwidth]{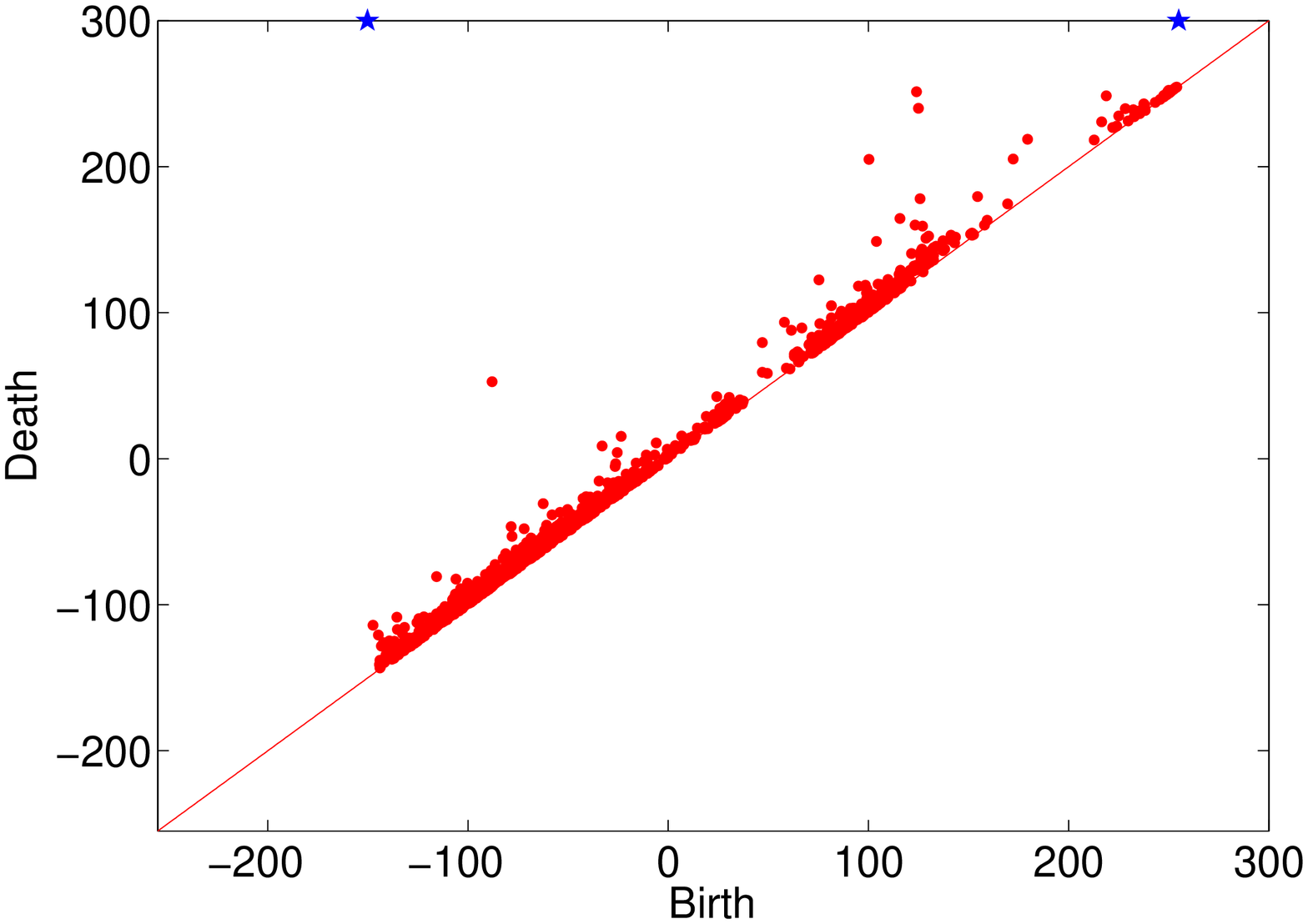}
	\label{fig:dgm-sph-step1}
	}
	\subfigure[Step 10]{
	\includegraphics[width=.3\textwidth]{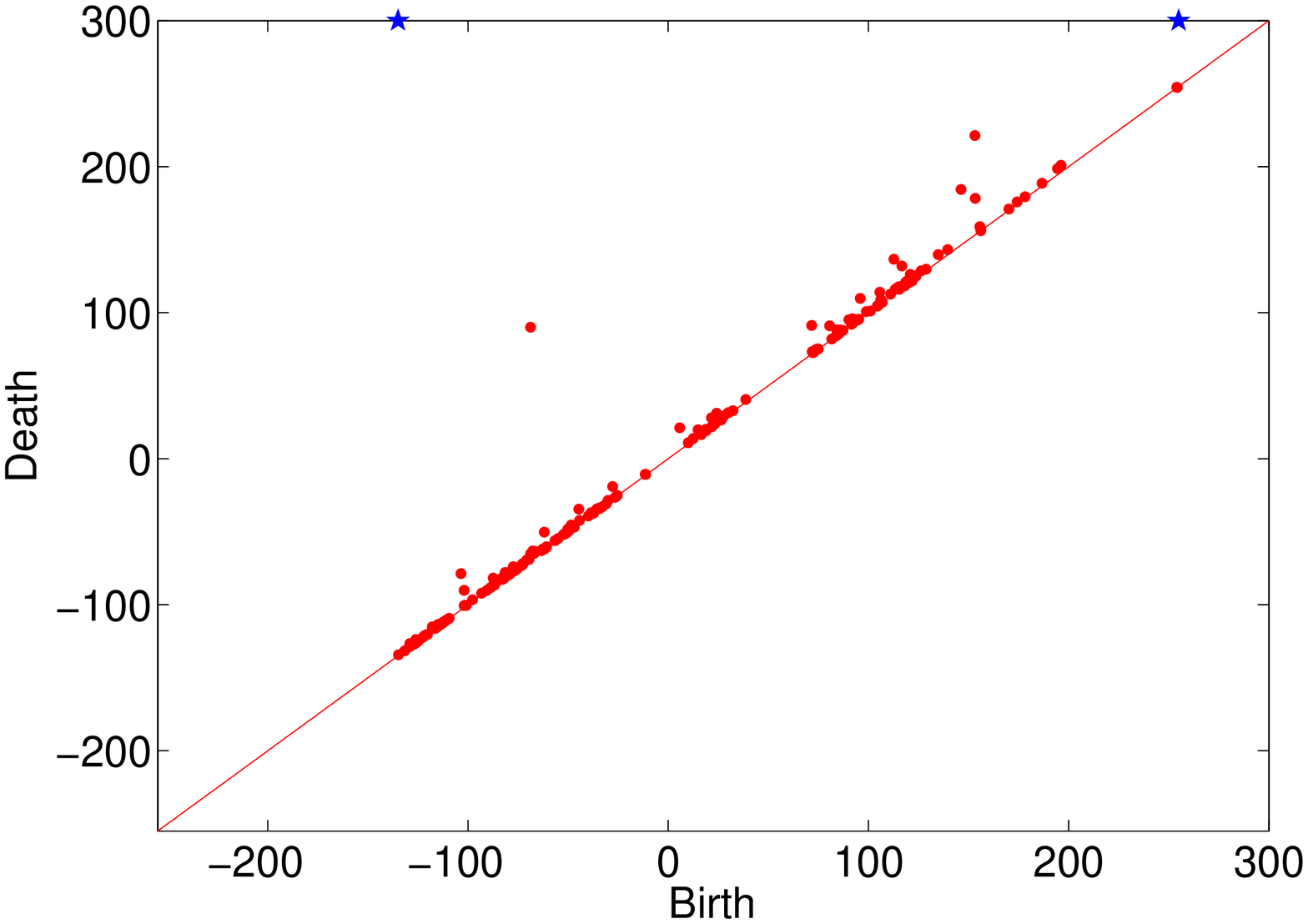}
	\label{fig:dgm-sph-step10}
	}
	\subfigure[Step 100]{
	\includegraphics[width=.3\textwidth]{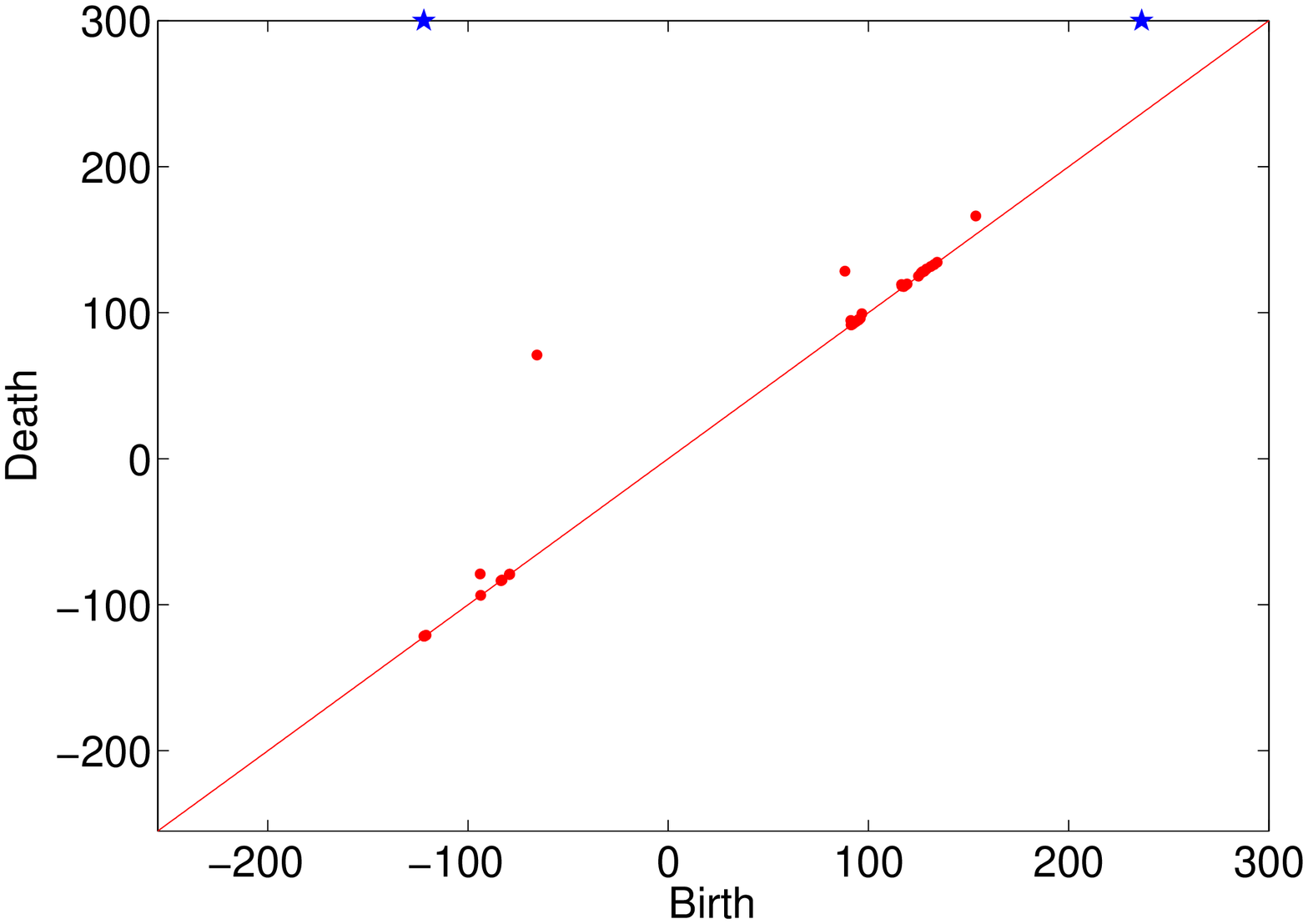}
	\label{fig:dgm-sph-step100}
        }
	\caption[Sphere Boundary]{Persistence diagrams for the heat equation using
                 spherical topology.
                }\label{fig:dgm-sph}
\end{figure}
In Figures \ref{fig-01:Sumi}-\ref{fig:dgm-sph}, we see several stages of the
heat equation homotopy using various topologies of the square.  The persistence 
diagrams shown are combined diagrams for all dimensions. 
\begin{figure}[hbt]
	\centering
	\subfigure[Klein Bottle, Dimension 0]{
	\includegraphics[width=.45\textwidth]{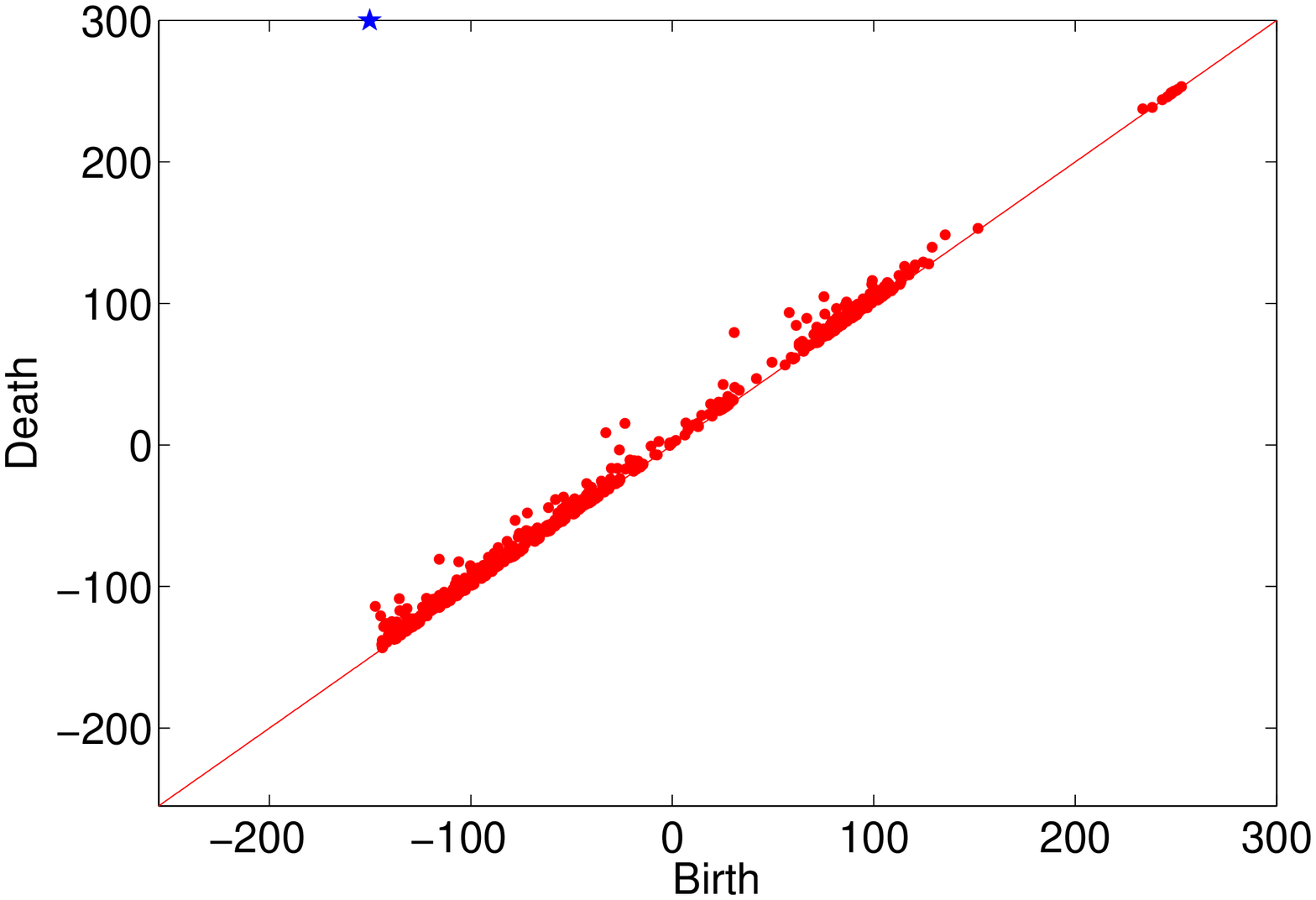}
	}
	\subfigure[Klein Bottle, Dimension 1]{
	\includegraphics[width=.45\textwidth]{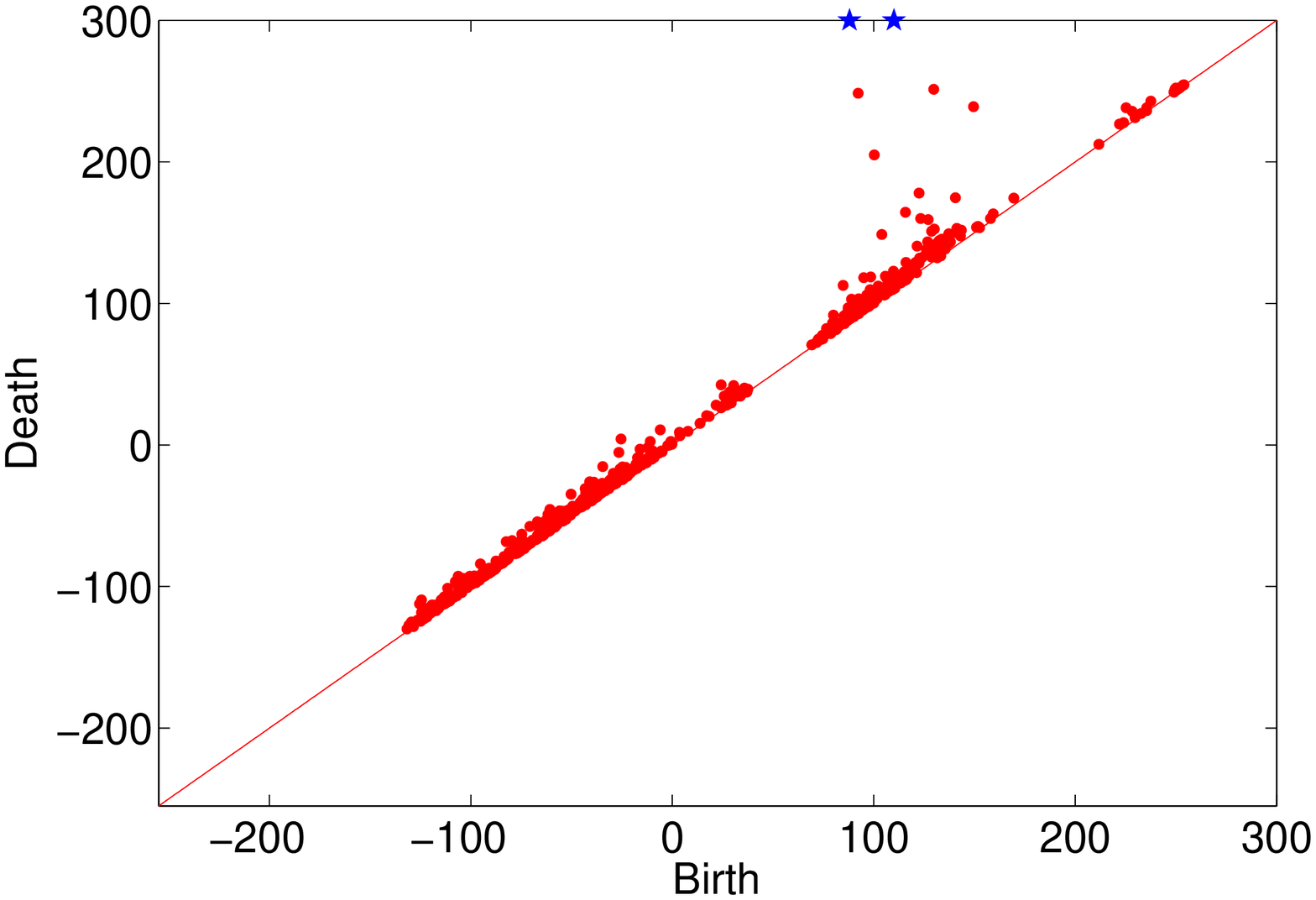}
	}
        \subfigure[Sphere, Dimension 0]{
	\includegraphics[width=.45\textwidth]{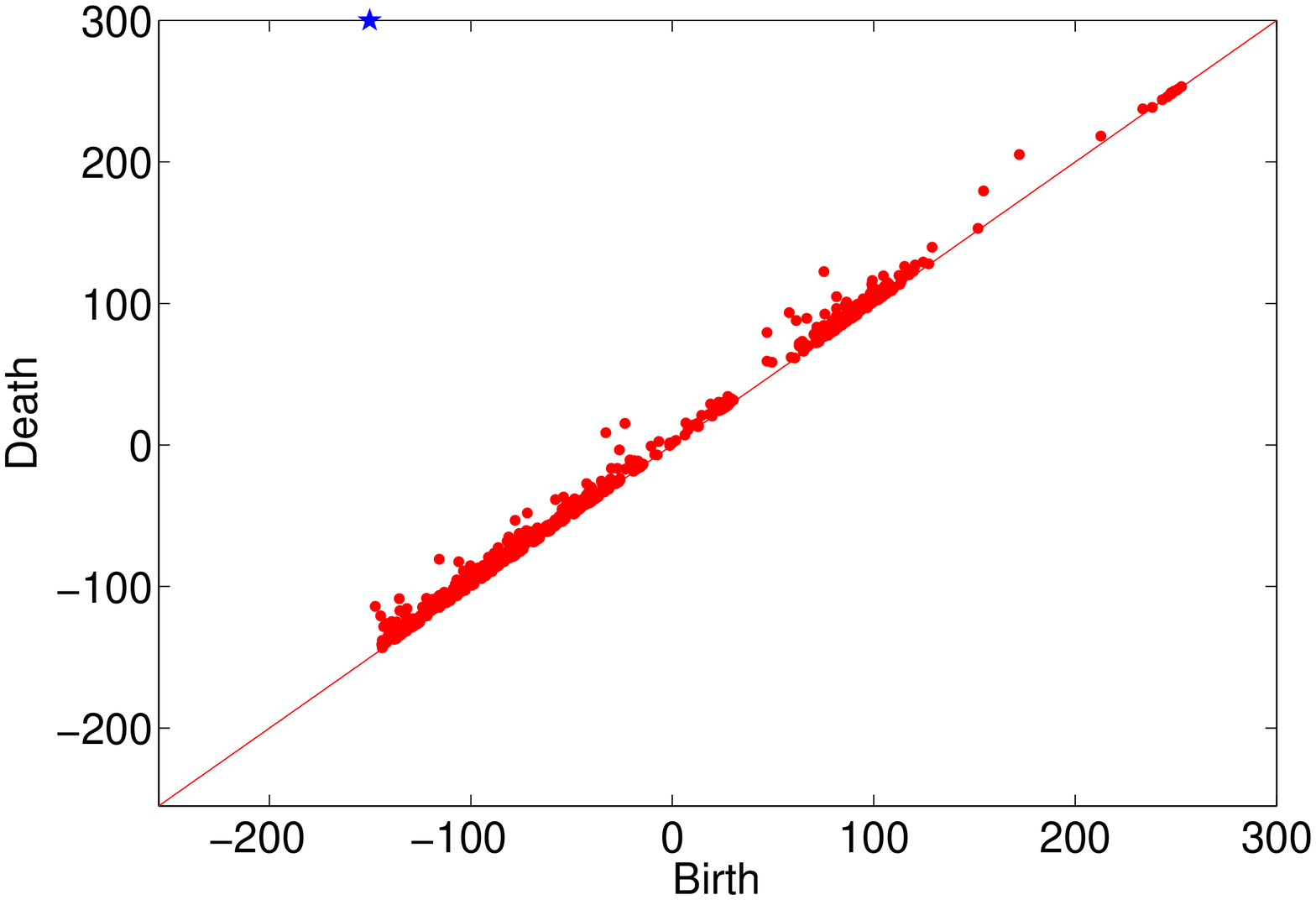}
	}
	\subfigure[Sphere, Dimension 1]{
	\includegraphics[width=.45\textwidth]{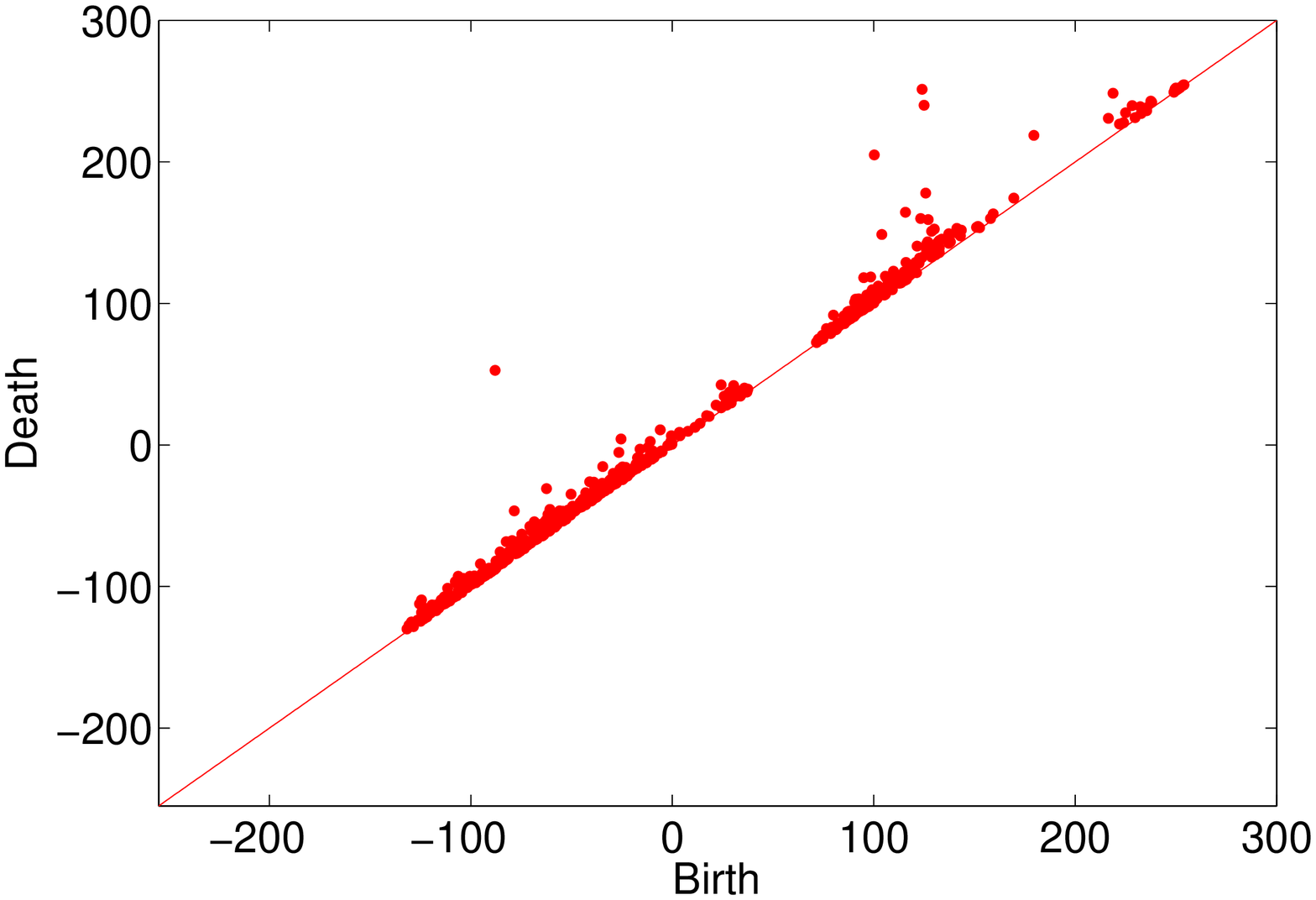}
	}
 \caption{In these figures, we separate the persistence diagrams from step one of the homotopy
          using the Klein bottle and the sphere topologies.  The diagrams for combined dimensions are
          \figref{fig:dgm-kb}(a) and \figref{fig:dgm-sph}(a).}
 \label{fig:splitDimensions}
\end{figure}
In \figref{fig:splitDimensions}, we look at the diagram for the first step
of the homotopy, separated for dimensions $p=0$ and 
$p=1$, under the Klein bottle and sphere topologies.  We note that the points
with the highest persistence appear in the dimension one persistence diagram.
This is a property that remains true as the homotopy progresses.

\subsection{Analyzing Different Topologies}
\begin{figure}[ht]
 \vspace*{0.1in}
 \centering
 \centerline{\epsfig{figure=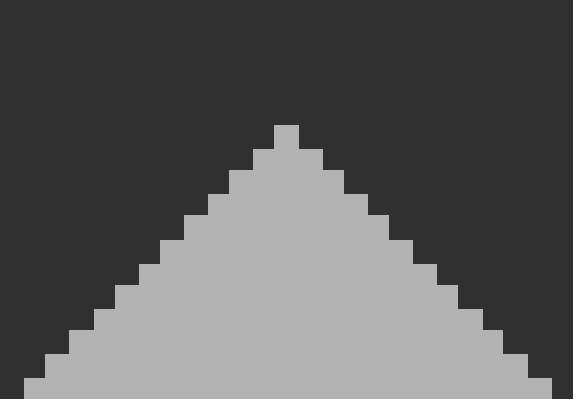, height=1in}}
 \caption[Triangle on Boundary]{
                 A square with a high-valued (light-colored)
                 region on the boundary.
                }\label{fig:tri-bdry}
\end{figure}
In the topologies without a boundary (torus, Klein bottle, and sphere), a new
feature is created with a relatively high persistence.  For example, if we 
start with a triangular region of high values against a border, as in 
\figref{fig:tri-bdry}, then there exists a $1$-cycle in the sublevel sets. 
Note, however, that there is not a $1-$cycle in any sublevel set of the square
topology.  Since we have created the four different topologies by gluing the 
edges of the square together in various ways, different behaviors along these
edges can be expected. We keep this difference in mind as we continue to look 
for commonalities and other differences caused by the adopted topology.

The persistence diagrams for the torus and the Klein
bottle topologies behave similarly.  In most of the graphs in this section, the
curves of the torus and the Klein bottle are usually parallel.  This is an 
indication that orientability has little effect on the heat equation homotopy.  
This does not come as a not a surprise, as we did not
use the orientation when computing the heat equation.

\subsection{Duration of Vines}
\begin{figure}[hbt]
 \vspace*{0.1in}
 \centering
 \centerline{\epsfig{figure=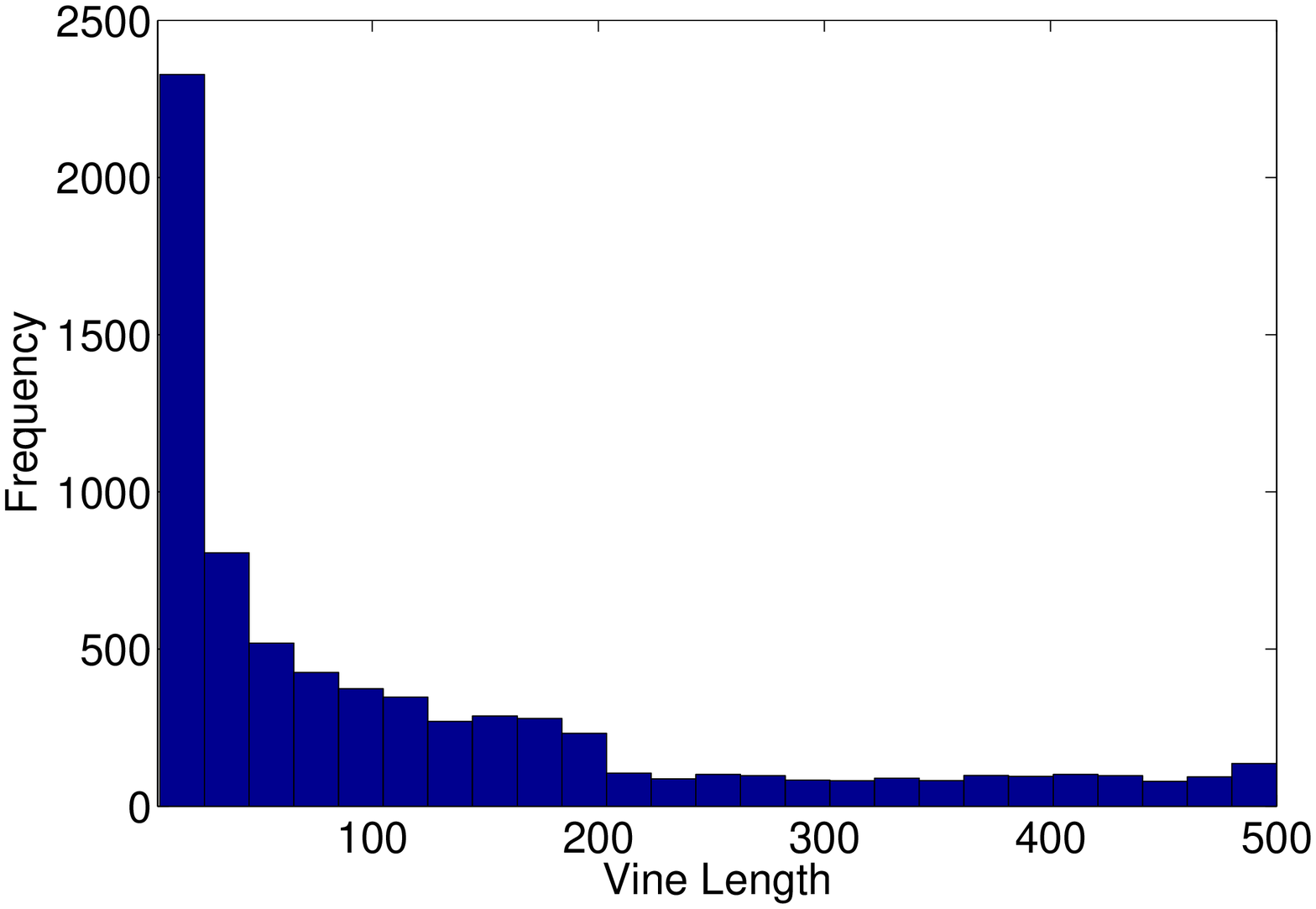,height=2.5in}}
 \caption{The distribution of the length of vines in the vineyard for Dgm$_0(u_t)$
          under the square topology.}
 \label{fig:vine_lengths}
\end{figure}
\begin{table}[h!b!p!]
\caption{Vine Length Statistics for Dgm$_0(u_t)$}
\begin{center} 
  \begin{tabular}{ |l | l | l | l | l | }
    \hline
          & mean & median & mode  & s.d.   \\ \hline\hline
   square & 63.2 & 15     & 6     & 132.8  \\ \hline
   sphere & 69.4 & 25     & 6     & 143.8  \\ \hline
   torus  & 79.8 & 26     & 500   & 105.1  \\ \hline
   Klein  & 72.5 & 21     & 500   & 101.4  \\ 
    \hline
  \end{tabular}
\end{center}\label{table:dgm0}
\end{table}
\begin{table}[h!b!p!]
\caption{Vine Length Statistics for Dgm$_1(u_t)$}
\begin{center} 
  \begin{tabular}{ |l | l | l | l | l | }
    \hline
          & mean  & median & mode & s.d.  \\ \hline\hline
   square & 119.6 & 65     & 6    & 97.7  \\ \hline
   sphere & 122.5 & 56     & 6    & 94.2  \\ \hline
   torus  & 122.5 & 70     & 6    & 137.1 \\ \hline
   Klein  & 126.2 & 58     & 6    & 151.3 \\ 
    \hline
  \end{tabular}
\end{center}\label{table:dgm1}
\end{table}
In \figref{fig:vine_lengths}, we see the distribution of the vine lengths in 
the vineyard Dgm$_0(u_t)$.  The histogram is skewed right, since the mean is 
greater than the median.  Table \ref{table:dgm0} confirms this observation.  
The same pattern is in fact observed under different topologies.  When 
comparing the vine lengths of any two topologies, the Kolmogorov-Smirnov test 
for statistical difference with $\alpha=.05$ fails to reject the null 
hypothesis that the distributions are the same.  On the other hand, if we 
remove the short-lived vines, then we start to see that comparing the Klein
bottle and sphere topologies results in the rejection of the null hypothesis.   
However, this is not a strong enough indication that these distributions are 
different.  The details of this statistical method are out of the scope of 
this paper, but can be found in ~\cite{massey1951kolmogorov}.

From these observations we can conclude that all topologies display a similar
distribution of the length of vines, with many of the vines being short-lived.  
In addition, under the torus and Klein bottle topologies, a large number
of vines span the entire vineyard.

\subsection{Monitoring Total Persistence}
The degree $q$ total persistence is the sum  of the $q^{th}$ powers of 
persistence over all points in the persistence diagram.
\begin{figure}[hbt]
 \vspace*{0.1in}
 \centering
 \centerline{\epsfig{figure=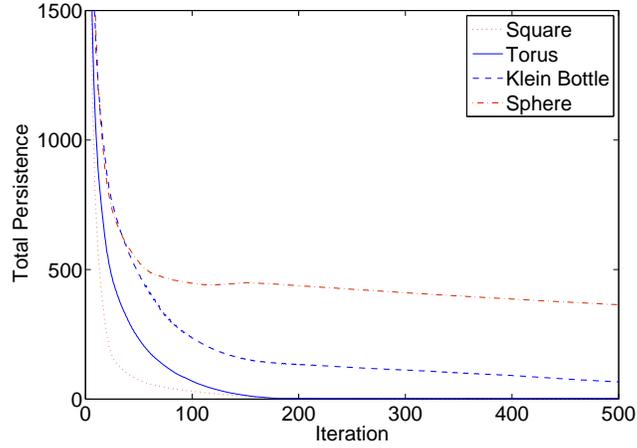,height=2.5in}}
 \caption{Total persistence of degree one.}
 \label{fig:totalPersis}
\end{figure}
\begin{figure}[hbt]
	\centering
	\subfigure[Degree 2]{
	\includegraphics[width=.45\textwidth]{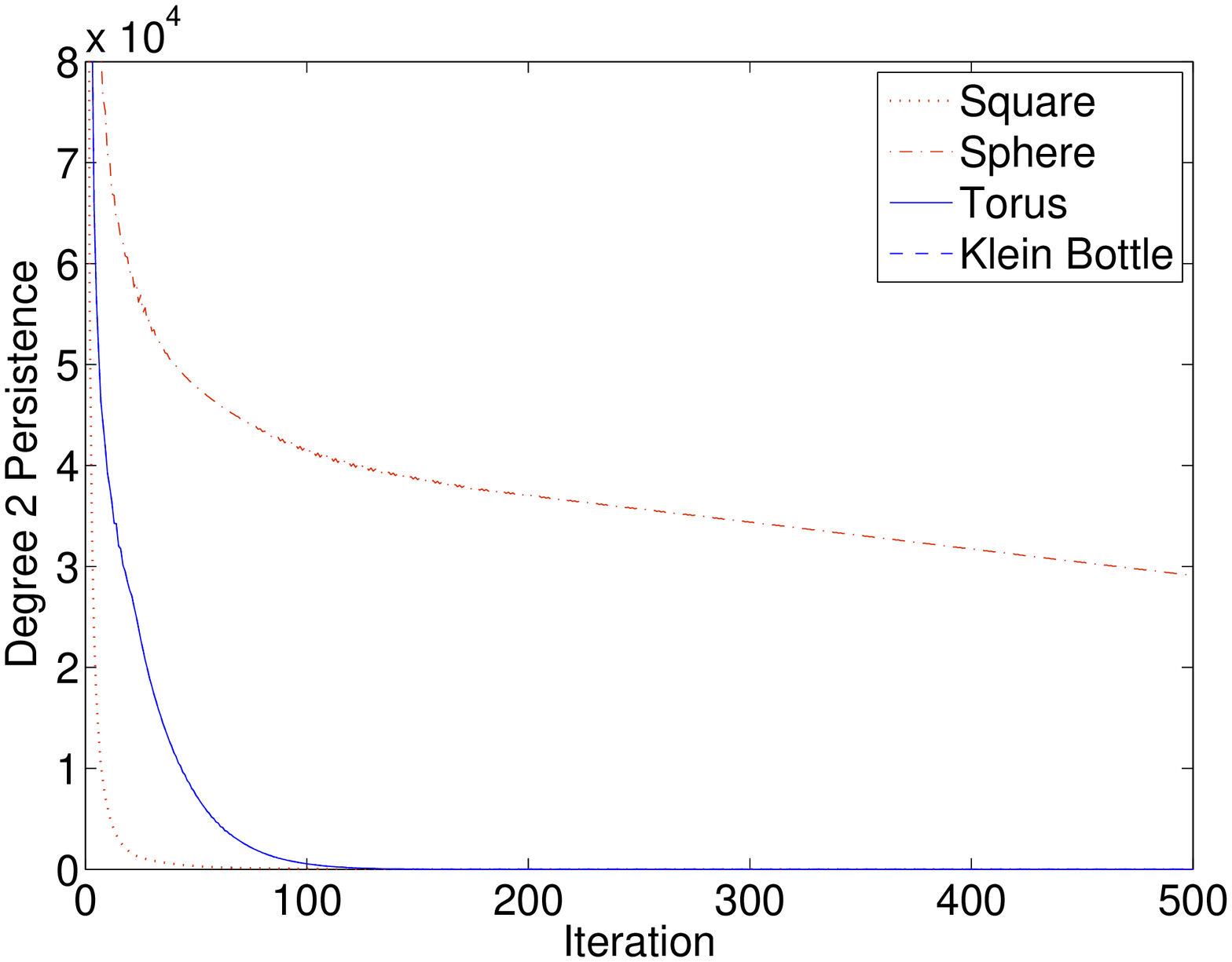}
	}
	\subfigure[Degree 3]{
	\includegraphics[width=.45\textwidth]{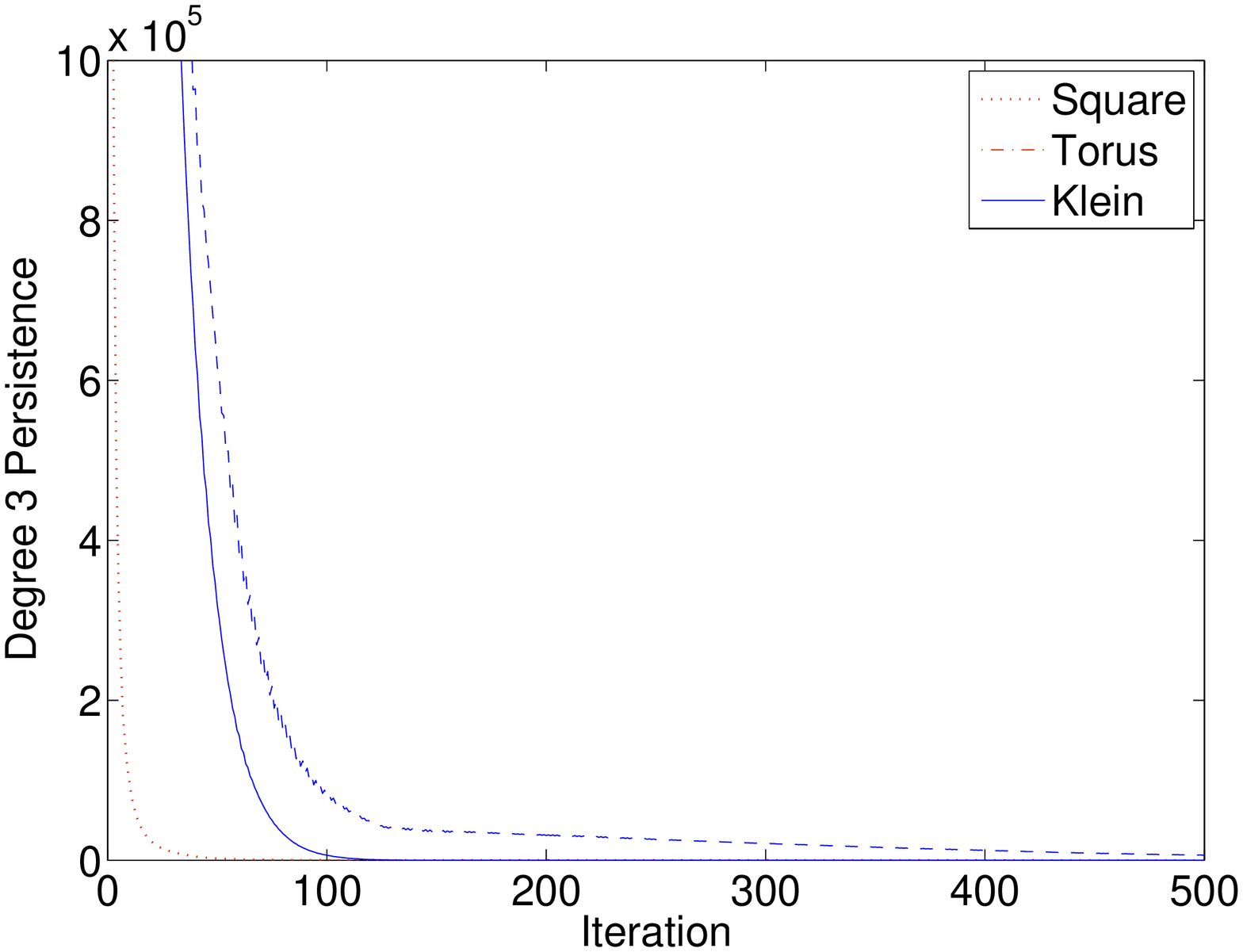}
	}
 \caption{The declining total persistence of degrees two and three.  In the graph for
          the total persistence of degree three, we omit the sphere topology.  Relative
          to the other topologies, the values were very high.}
 \label{fig:totalPersis23}
\end{figure}
In \figref{fig:totalPersis}, we have the graph of the degree one total persistence
versus the iteration.  Although the total persistence rapidly 
deteriorates initially, the decay slows down around iteration $100$.
-
-
We notice here that the Klein bottle and the
torus have an end behavior different than that of the sphere and the square,
in that we do not see the total persistence approaching zero
after $500$ steps of the heat equation.  
-
In these figures, the sphere behaves
radically different than the other three topologies.

\subsection{Mean Absolute Change}
\begin{figure}[hbt]
 \vspace*{0.1in}
 \centering
 \centerline{\epsfig{figure=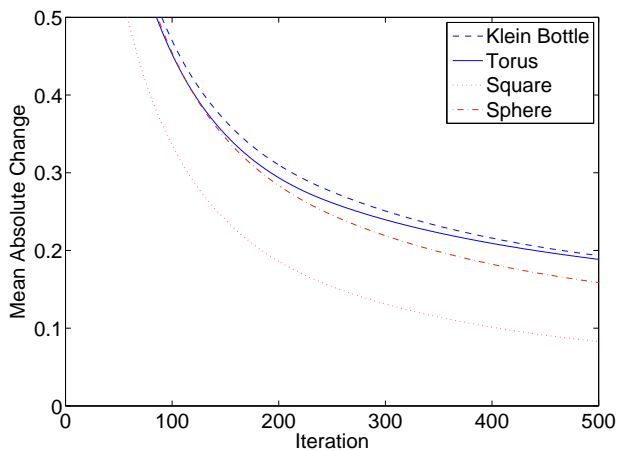,height=2.5in}}
 \caption{The mean absolute change between steps of the heat equation homotopy.}
 \label{fig:meanChange}
\end{figure}
\figref{fig:meanChange} shows that the mean absolute value of the change 
between steps
of the homotopy decreases rapidly at first, then slowly.  Recall that the values of the 
mesh points range from $-255$ to $255$.  At the first step, the mean
absolute change is between $10$ and $15$ (not shown in \figref{fig:meanChange}), 
which is only a $2\%-3\%$ initial change.  The value, 
however, remains above $0.1$ in all cases except for the square
topology.  Given the nature of the heat equation,
we expect the values to decrease with respect to time.
The small values for mean absolute change are in part due to the initial values of
$u_0$.  The values of one vertex does not differ by a large amount from the values
of its neighbors.  
We know
that the iterative method chosen is slow to converge; however, this
graph allows us to gauge how little change is occurring at each step of the
homotopy.

\subsection{Counting Transpositions}
\begin{figure}[hbt]
	\centering
	\subfigure[Total]{
	\includegraphics[width=.45\textwidth]{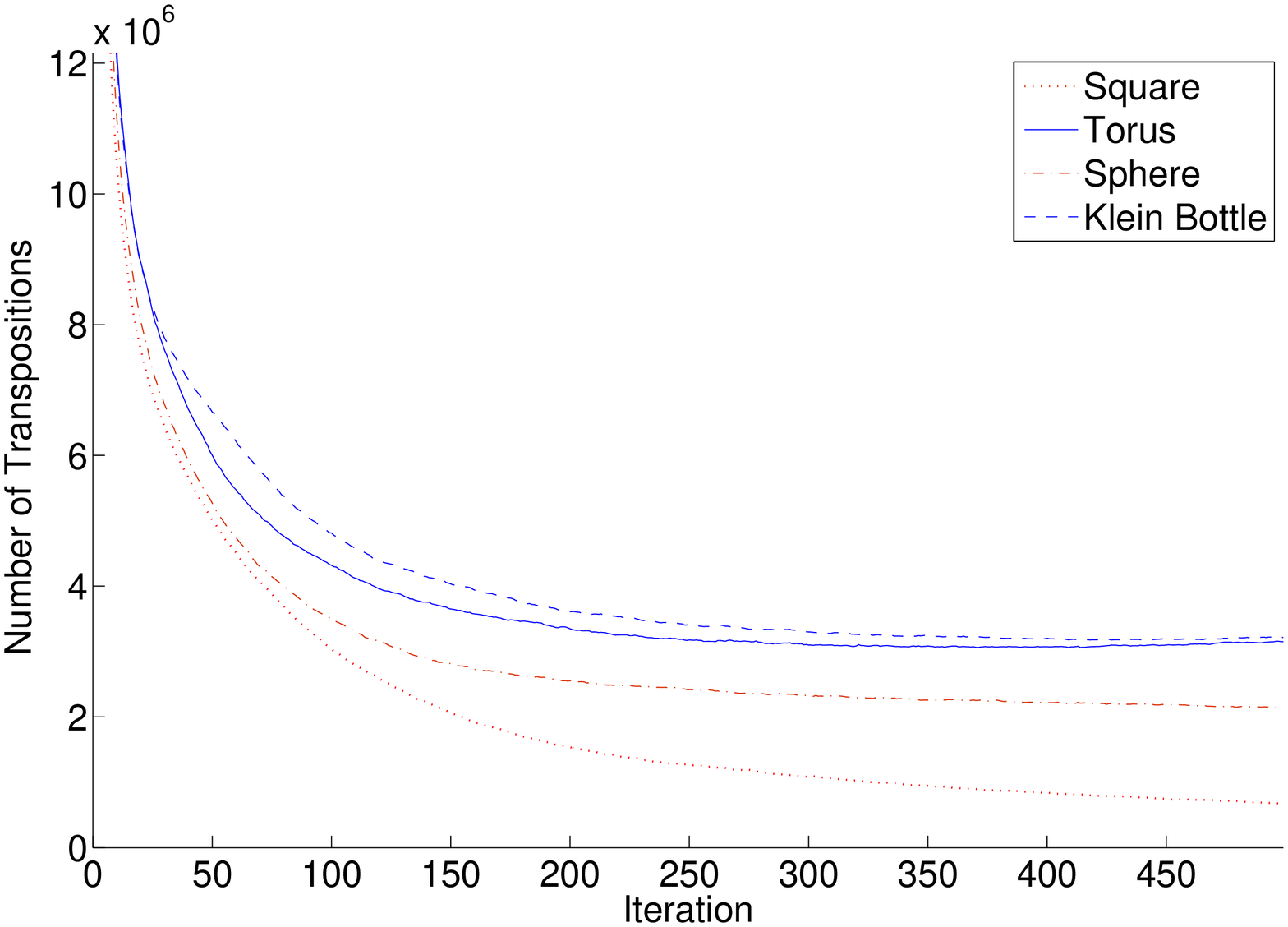}
	\label{fig:a}
	}
	\subfigure[Type 1 Pair Swap]{
	\includegraphics[width=.45\textwidth]{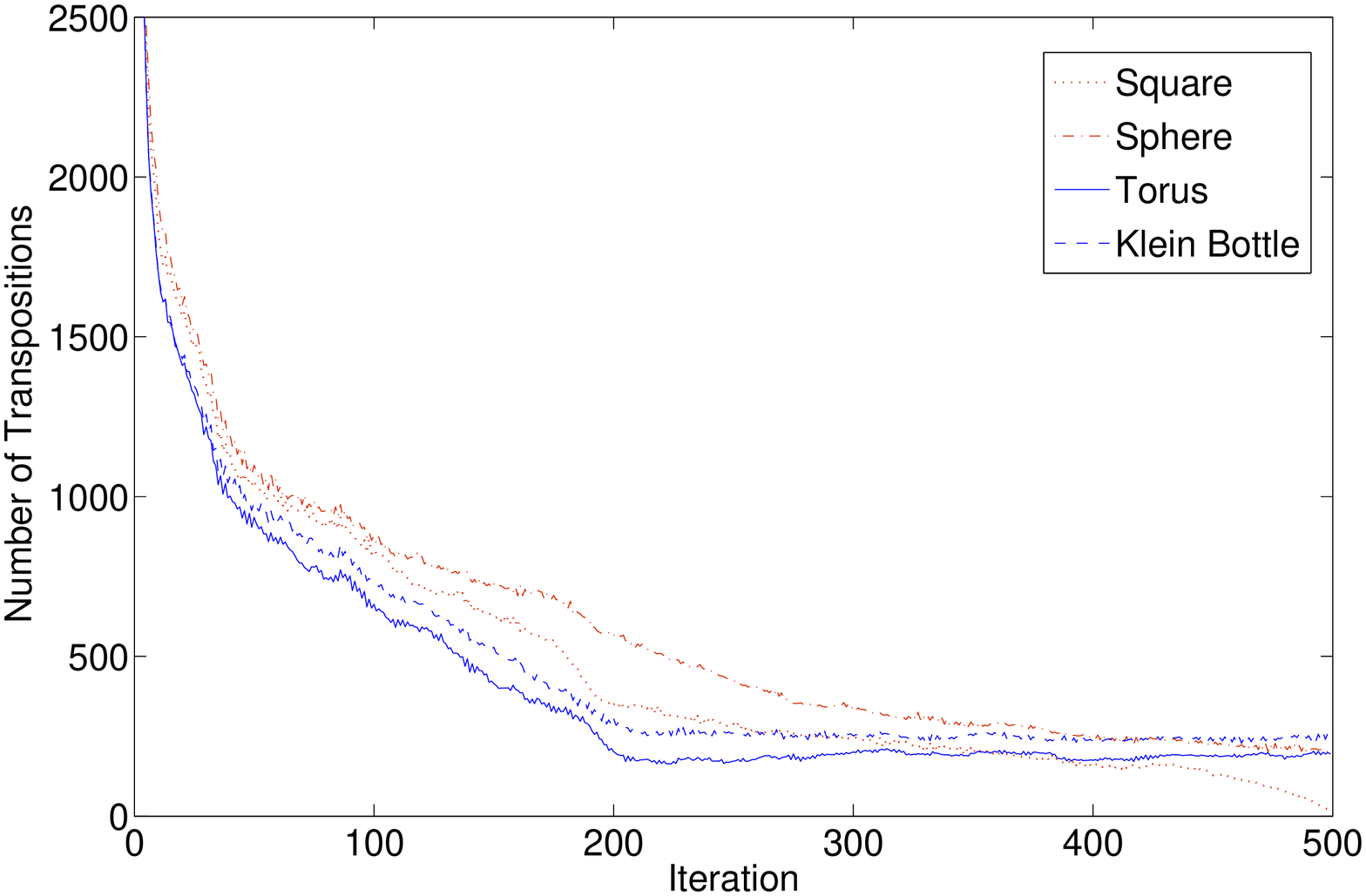}
	\label{fig:b}
	}
	\subfigure[Type 2 Pair Swap]{
	\includegraphics[width=.45\textwidth]{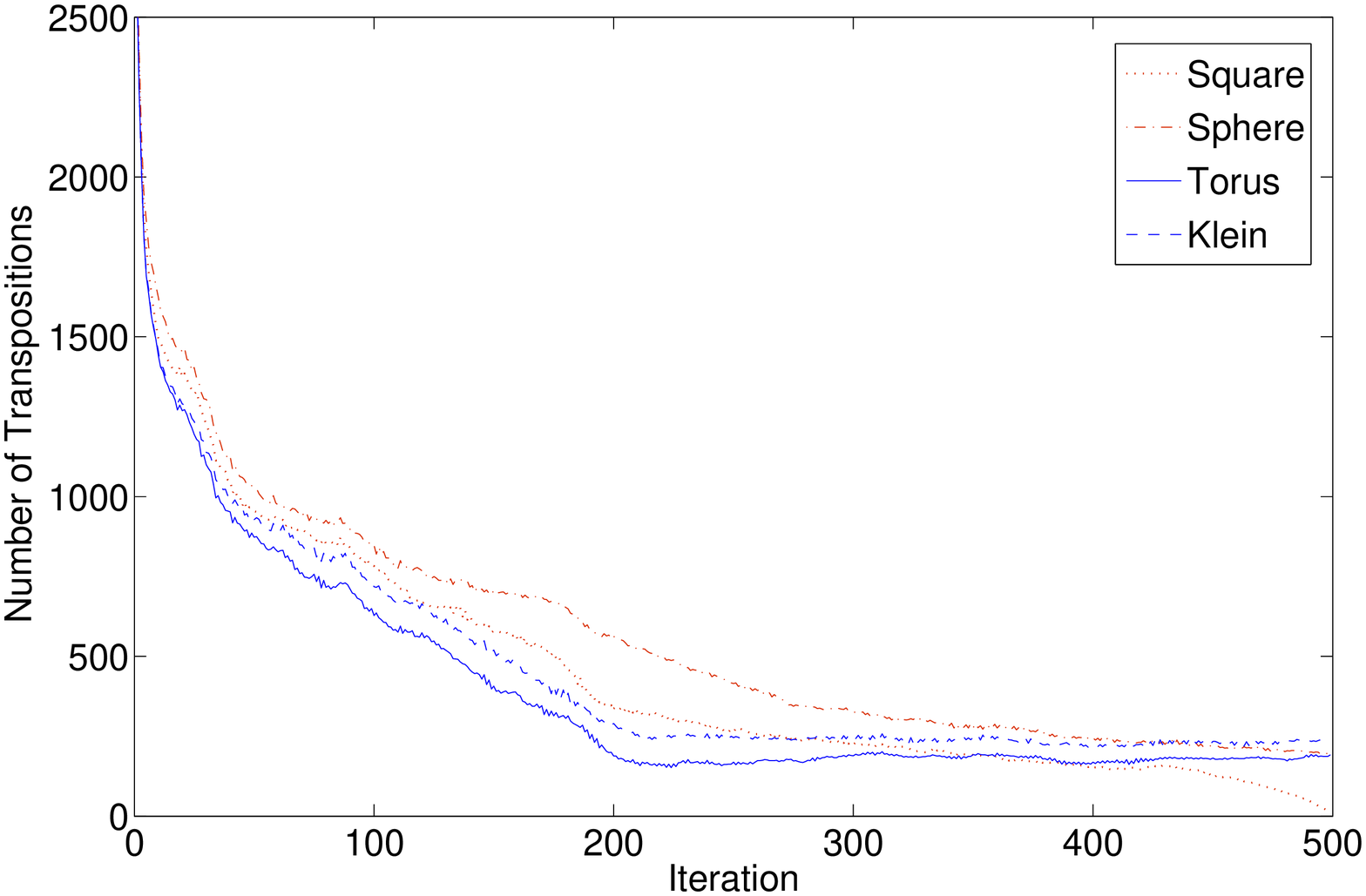}
	\label{fig:ax}
	}
	\subfigure[Type 3 Pair Swap]{
	\includegraphics[width=.45\textwidth]{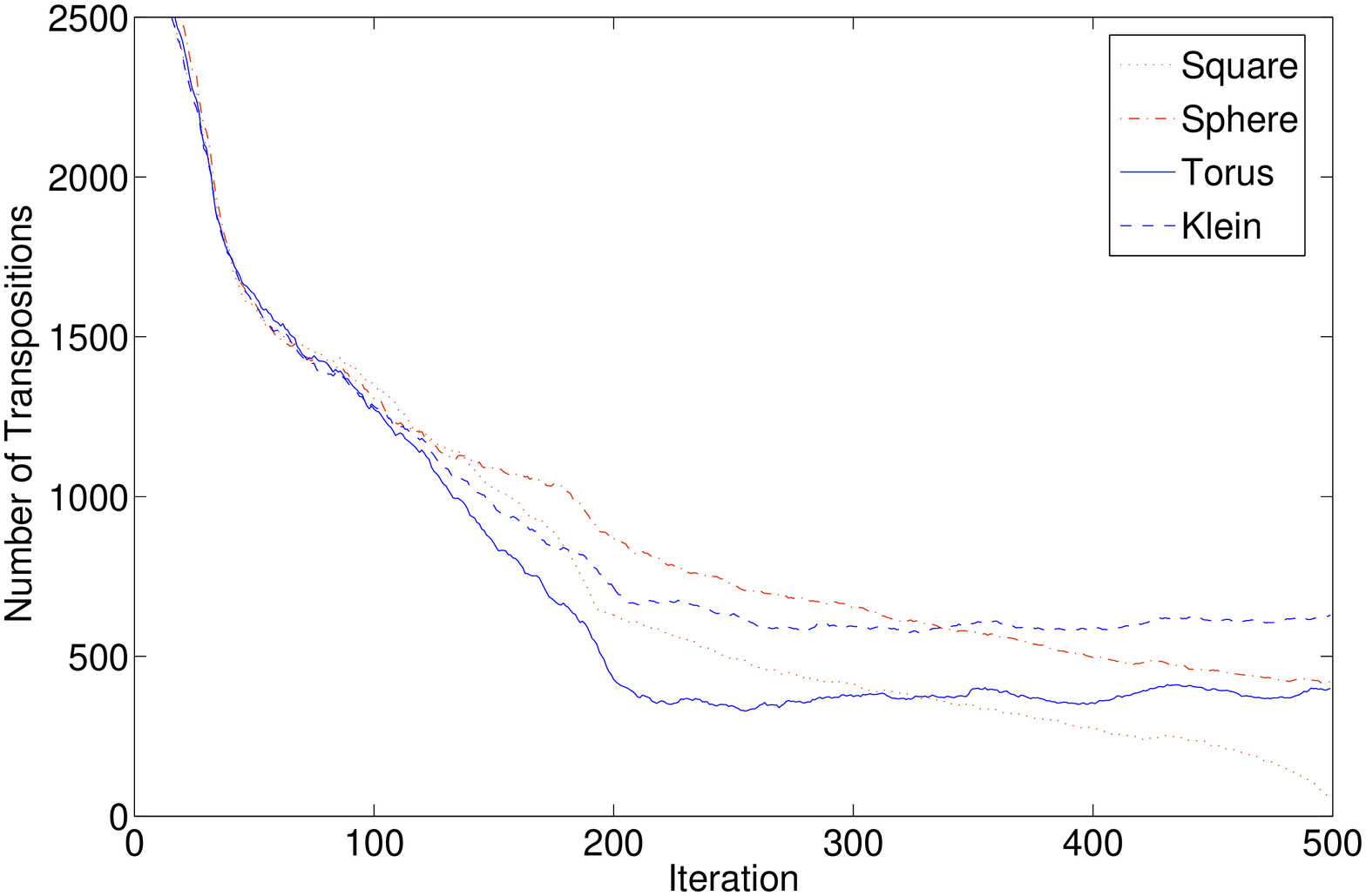}
	\label{fig:bx}
	}
 \caption{The number of transpositions between steps of the heat equation
          homotopy.  The top left shows the total number of transpositions and
          the remaining three figures are restricted to the three types of
          transpositions that result in pairing swaps.}
 \label{fig:numTrans}
\end{figure}

The total number transpositions between steps of the heat equation versus 
time is shown in \figref{fig:numTrans}(a).  All four topologies follow a 
decreasing pattern that levels off, with the Klein bottle and the torus behaving
distinctively different from the square and the sphere.  

We remark on several notable observations from these diagrams.  
The total number of transpositions is 
still significant when the heat equation algorithm reaches
the halting condition.  At the last step, the square topology makes over
400,000 total transpositions.  The other three topologies make even more
transpositions.

In \figref{fig:numTrans} (b),(c), and (d) to see a different pattern for 
the number of the switches resulting from the transpositions.
Under the square topology, the number of pair swaps of types 1,
2, and 3 is down to the double digits after $500$ iterations of the heat 
equation.
We notice that Figures (b) and (c) are almost identical.  This is not a 
surprising observation, since pairing swaps 1 and 2 (described in 
\secref{ss:transpose}) are symmetric cases of two births of the same dimension
or two deaths of the same dimension being transposed, resulting in a pair swap
in the diagram.

Three other patterns are worth noting in the graphs of \figref{fig:numTrans}.  
First, we see that
the number of transpositions (of all kinds) rapidly decreases until 
iteration ~$25$.  At this point, the four topologies show different behaviors.
Second, after iteration ~$200$, the torus and the Klein bottle topologies have
both leveled off to a constant function.
Finally, we notice that adding graphs (b), (c), and (d) does not result
in a graph that looks like (a).  Thus, a significant number of the
transpositions made are those that do not result in a pairing swap.
Proportionally, this occurs more often in the torus and Klein bottle topologies
than in the square and the sphere topologies.

\begin{figure}[hbt]
 \vspace*{0.1in}
 \centering
 \centerline{\epsfig{figure=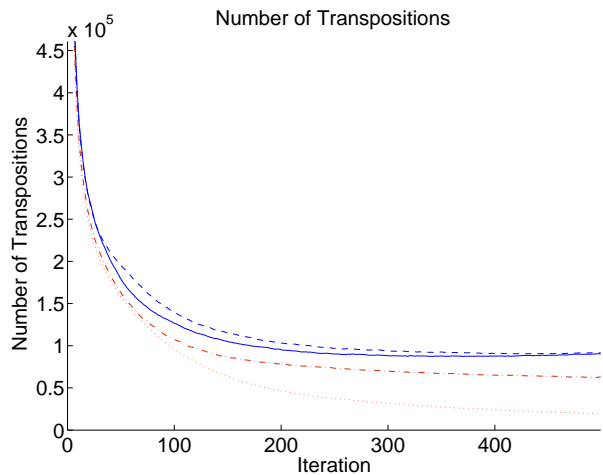,height=2.5in}}
 \caption{The number of transpositions between vertices versus the iteration
          of the heat equation.}
 \label{fig:vertexTranspositions}
\end{figure}
In \figref{fig:vertexTranspositions}, we restrict our counts to the number 
transpositions of vertices only.  The
values at the vertices dictates the values of the edges and the faces.  The
pattern that this graph follows is similar to 
\figref{fig:numTrans}(a), the graph for the number of transpositions 
generated from all of the simplices  Thus, is seems that the behavior of the vertex-vertex 
transpositions is proportional to the behavior of simplex-simplex 
transpositions.

\section{Conclusion}\label{sec:conclusion}
The objective of this project was to find a new way of measuring the distance
between two functions.  We develop a homotopy of functions based on the
heat equation and investigated the behavior of the vineyard resulting
from this homotopy.  We computed this homotopy using four topologies,
and found some trends that differed from our expectation.  For example,
the heat equation on elliptic surfaces is known to converge more quickly
than in Euclidean space.  Yet, we found the behavior of the sphere
topology to contradict this fact.  We believe that this contradiction
arises from the discretization of the sphere.

We began an investigation of the behavior of the heat
equation homotopy; however, many directions still remain for where we can 
continue.  We would like to explore if the observations made in 
\secref{sec:results} were specific to initial function $u_0$, or if we are 
observing behaviors that arise from the topologies used in the heat equation.
Moreover, we would like to formally compare the matching we obtain with the 
bottleneck and Wasserstein matchings.

\bibliographystyle{acm}
\bibliography{myBib}

\end{document}